\documentclass[10pt]{article}
\usepackage{geometry} 
\usepackage[hang,bf]{caption}
\usepackage{setspace}
\usepackage[parfill]{parskip}    
\usepackage{times}
\usepackage{graphicx}
\usepackage{amssymb,amsmath,amsbsy}
\usepackage{cancel}
\usepackage{epstopdf}
\usepackage{threeparttable}
\usepackage{url}
\usepackage{epsfig}
\usepackage{cite}
\usepackage[export]{adjustbox}
 \usepackage{url}
 \PassOptionsToPackage{hyphens}{url}\usepackage{hyperref}

\begin{document}
\title{\textbf{Radon Mitigation Applications at the Laboratorio Subterr\'aneo de Canfranc (LSC)\\\small{Published in MDPI, Special Issue ``Studying the Universe from Spain'' \url{https://doi.org/10.3390/universe8020112}}}}


\author{J. P\'erez-P\'erez$^{1,4}$ \footnote{Corresponding author. Current address: Affiliation 4. Correspondence: javier.perez.perez@uj.edu.pl or jperez@lsc-canfranc.es} , J.C. Amare$^{2}$, I.C.  Bandac $^{1}$, A. Bayo$^{1}$, S. Borjabad-S\'anchez $^{1}$, \\   J.M. Calvo-Mozota $^{1,6,7}$, L. Cid-Barrio $^{1}$, R. Hern\'andez-Antol\'in  $^{1}$, B. Hern\'andez-Molinero  $^{1}$, \\ P. Novella$^{3}$, K. Pelczar$^{4}$, C. Pe\~na-Garay $^{1}$,  B. Romeo$^{1,5}$, A. Ortiz de Sol\'orzano $^{2}$, M. Sorel$^{3}$, \\ J. Torrent$^{5}$, A. Us\'on$^{3}$, A. Wojna-Pelczar$^{4}$ and G. Zuzel$^{4}$.}

\date{}
\maketitle

\begin{center}
\small{
$^{1}$  Laboratorio Subterr\'aneo de Canfranc (LSC), Canfranc-Estaci\'on 22880, Spain\\
$^{2}$  Centro de Astropart\'iculas y F\'isica de Altas Energ\'ias (CAPA), Universidad de Zaragoza, E-50009 Zaragoza, Spain\\
$^{3}$  Instituto de Física Corpuscular (IFIC), CSIC \& Universtitat de Val\`encia, Calle Catedrático José Beltrán, 2, Paterna, E-46980\\
$^{4}$  M. Smoluchowski Institute of Physics, Jagiellonian University, ul. \L{}ojasiewicza 11, 30-348 Krak\'ow, Poland\\
$^{5}$  Donostia International Physics Center, (DIPC), Paseo Manuel Lardizabal, 4, Donostia-San Sebastián, E-20018, Spain\\
$^{6}$  Universidad Internacional de La Rioja, Avenida de la Paz, 137, 26006 Logroño, La Rioja, Spain\\
$^{7}$  Universidad Pública de Navarra, Campus de Arrosadia, 31006 Pamplona, Spain}

\end{center}

\begin{abstract}\
The Laboratorio Subterr\'aneo de Canfranc (LSC) is the Spanish national hub for low radioactivity techniques and the associated scientific and technological applications. The concentration of the airborne radon is a major component of the radioactive budget in the neighborhood of the detectors. The LSC hosts a Radon Abatement System, which delivers a radon suppressed air with $1.1 \pm 0.2$ mBq/m$^3$ of $^{222}$Rn. The radon content in the air is continuously monitored with an Electrostatic Radon Monitor. Measurements with the double beta decay demonstrators NEXT-NEW and CROSS and the gamma HPGe detectors show the important reduction of the radioactive background due to the purified air in the vicinity of the detectors. We also discuss the use of this facility in the LSC current program which includes NEXT-100, low background biology experiments and radiopure copper electroformation equipment placed in the radon-free clean room.
\end{abstract}


\section{INTRODUCTION}\label{sec:1}
\begin{figure}[htbp] 
   \centering
   \includegraphics[width=13.5cm]{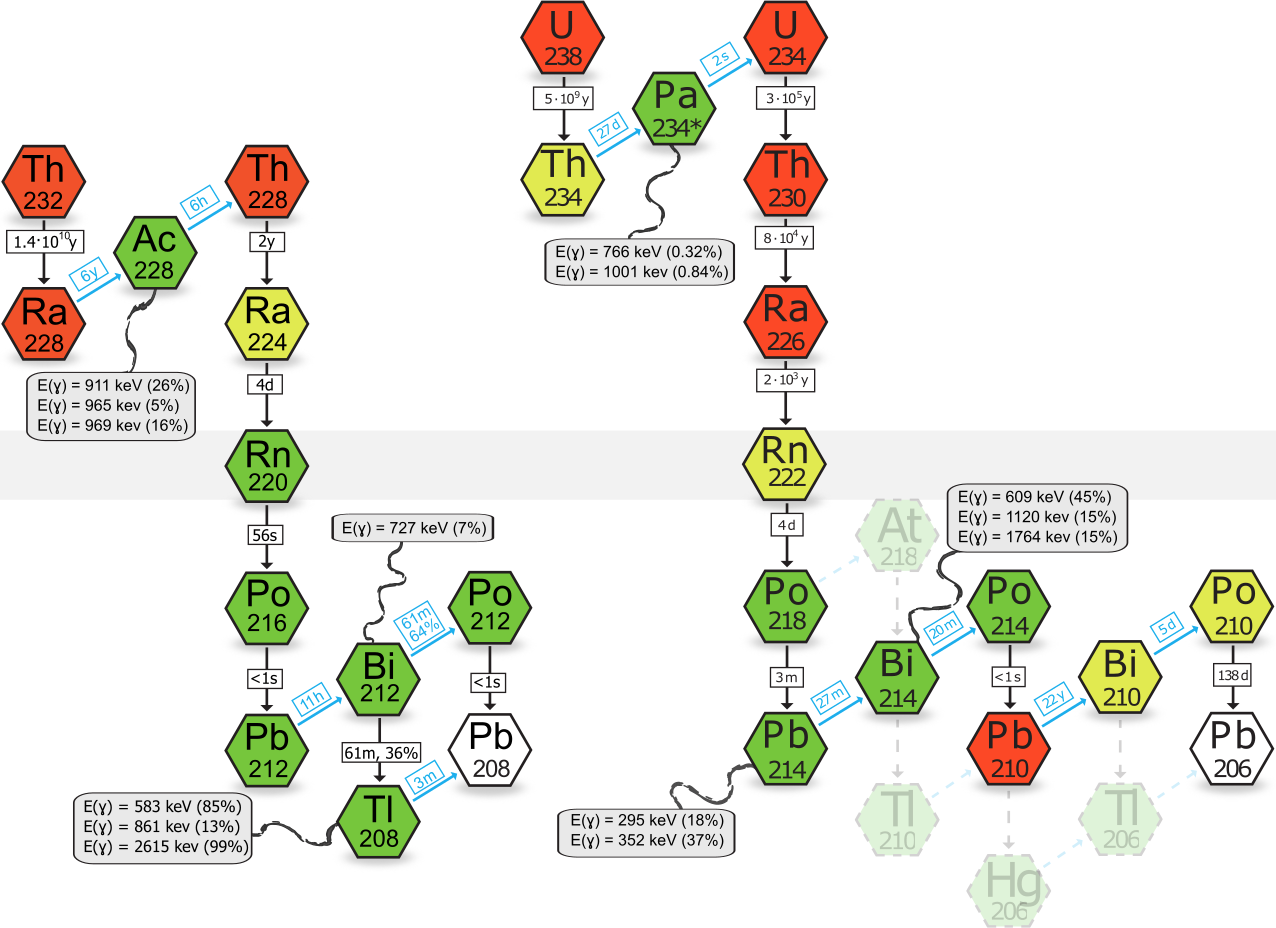} 
   \caption{\emph{Radioactive chains. Left: $^{232}$Th decay chain; Right: $^{238}$U decay chain. In the grey boxes the most important gamma peaks energies and decay probabilities of the corresponding isotopes are presented.}}
   \label{fig:rad_chains}
\end{figure}

Underground laboratories are the best places to explore science and develop technology in ultra-low radioactive background \cite{ref-journal50}. These laboratories have a rock shielding that blocks most of the cosmic rays, reducing the external background of the experiments several orders of magnitude. On the other hand, the surrounding rocks can have large concentrations of $^{232}$Th, $^{238}$U, and $^{235}$U. These isotopes start the natural decay series that involve several radioisotopes and the emission of alpha, beta and gamma particles (see Figure \ref{fig:rad_chains}) \cite{ref-journal1}.

One of the elements of these decay chains is radon, a noble gas that can be emanated from the materials and increases laboratory air radioactivity due to its unstable isotopes. (e.g., $^{222}$Rn from $^{238}$U, $^{219}$Rn from $^{235}$U and $^{220}$Rn from $^{232}$Th). The most important radioactive background source among these radon isotopes is $^{222}$Rn from $^{238}$U chain, with the half-life of $3.82$ days. This half-life is long enough to allow this radioactive isotope to escape from the material's surface, which increases significantly the overall radioactive background. 

In Figure \ref{fig:rad_chains} we used four different colours to identify isotopes: green for isotopes that decay fast (hours, seconds or shorter), yellow for intermediate decay time (days or months), red for long half-life isotopes (years) and white for the stable isotopes (the end of the different chains). When the secular equilibrium is broken, every isotope in red initiates a sub-chain in equilibrium. Light colored radioisotopes and arrows show low probability decay paths. Black downwards arrows represent alpha decays, while blue up-diagonal ones, beta decays. Half-lifes of the isotope and decay probabilities of the branch are indicated next to the arrows. Finally, grey floating boxes contain the most relevant gamma energies emitted in the decays.


\section{LABORATORIO SUBTERR\'ANEO DE CANFRANC}\label{sec:2}
\begin{figure}[htbp] 
   \centering
   \includegraphics[width=15cm]{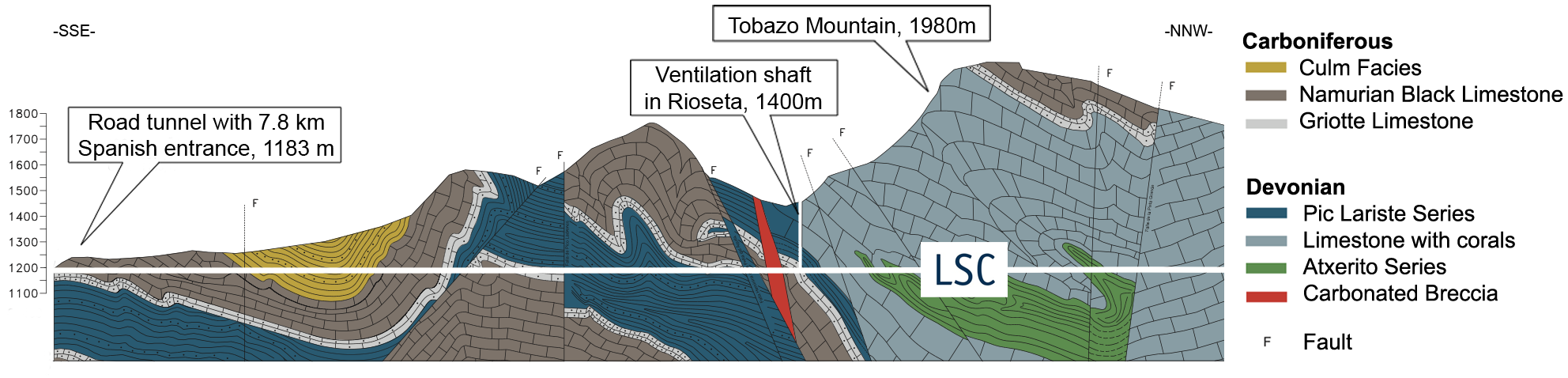} 
   \caption{\emph{Laboratorio Subterr\'aneo de Canfranc (LSC) - Longitudinal geological profile.}}
   \label{fig:lsc}
\end{figure}

Laboratorio Subterr\'aneo de Canfranc (LSC) \cite{ref-url} is located in the Spanish side of the Pyrenees Mountains, under the Tobazo peak. It is situated between the international Somport road tunnel (that connects Spain and France) and the old train tunnel (not in use). The laboratory has a rock shielding of 800 m  (Figure \ref{fig:lsc}), equivalent to approximately 2400 meter water equivalent (mwe), that suppresses the cosmic muon flux by almost 5 orders of magnitude.

The total area of the underground laboratory is about 1600 m$^{2}$, corresponding to a volume of about 10000 m$^{3}$. The laboratory is divided in four main areas: Hall A, Hall B, Hall C, Hall D and Hall E. Inside the laboratory several services are provided like a Clean Room (Class ISO 7 / Class 10000 and a Class ISO 6 / Class 1000 in part of the room), an Ultralow Background Service (ULBS) using Gamma Spectrometry with High Purity Germanium Detectors (HPGe) situated in Hall C, a mechanical workshop, the Radon Abatement System (RAS) and a liquid nitrogen production and distribution facility. 

The fresh air inside the laboratory comes from the Rioseta air entrance, located on surface at 1400 meters altitude. This air has to cross a 230 m vertical tunnel followed by a 700 m horizontal concrete tunnel excavated in the rock before being collected in a 100 m pipe that connects the tunnel and the laboratory. The air arrives to the Air Conditioning Room (Sala de Climatizaci\'on, SC) of the laboratory and finally it is distributed in all the rooms of the laboratory with steel pipes, with a flow of 25000 m$^{3}$/h. This quantity plus the return air vent is enough to exchange all the air of the laboratory several times per hour. 

\begin{figure}[htbp] 
   \centering
   \includegraphics[width=15cm]{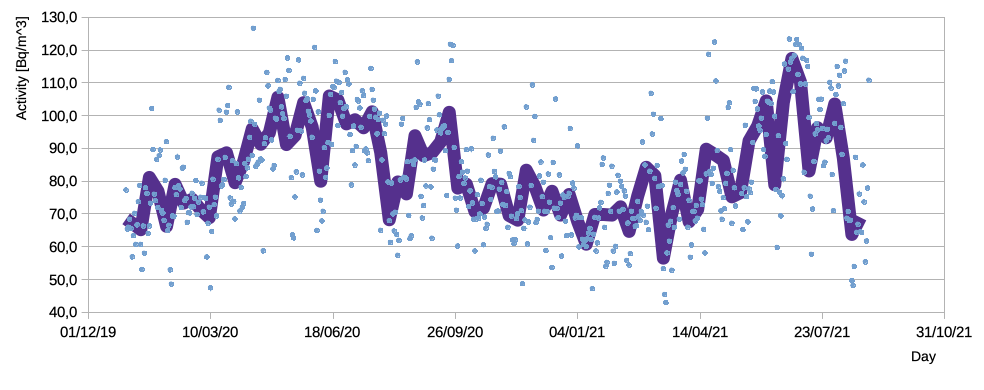} 
   \caption{\emph{Radon measurements inside Hall A. Light blue points show the daily measurements and the blue line shows the weekly mean value.}}
   \label{fig:RnHallA}
\end{figure}

The radon concentration in air is monitored with AlphaGuard detectors (AlphaGuard Saphimo P30  in Hall A, B and C; AlphaGuard Saphimo P2000 in Hall E). These detectors have a sensitivity of 1 CPM at 20 Bq/m$^{3}$, a lower limit of $\sim 2$ Bq/m$^{3}$, and a calibration error of $\pm 3\%$. The average  annual activity in the underground lab common area is ($69.0\pm 0.3$) Bq/m$^{3}$ with daily, seasonal and annual fluctuations (see Figure \ref{fig:RnHallA} for Rn measurements performed in Hall A). The values obtained in different experimental halls inside LSC performed at the same time are compatibles \cite{ref-journal4}. The radon emanation depends on several external factors including rain or snow melting because water increases radon diffusion from the material surfaces and its later emanation \cite{ref-journal1}, \cite{ref-journal51}, \cite{ref-journal52}, \cite{ref-journal53}, \cite{ref-journal54}.

A dedicated campaign of measurements was performed in the underground lab to identify the spatial distribution of the radon sources \cite{ref-journal3}. An AlphaGuard P30 detector was placed in several locations between the air entrance at Rioseta and the pipe that collects this air and guide it towards the Air Conditioning Room situated inside the underground laboratory. We observed an increase in the radon activity from about 20 Bq/m$^{3}$ at the entrance on surface (Rioseta) to approximately $\sim 70$ Bq/m$^{3}$ at the entrance of the air inside the laboratory. This increase is caused by the radon emanated by the concrete of the two tunnels from the Rioseta entrance to the laboratory.


\section{RADON ABATEMENT SYSTEM}\label{sec:3}

\begin{figure}[htpb] 
   \centering
   \includegraphics[width=12cm]{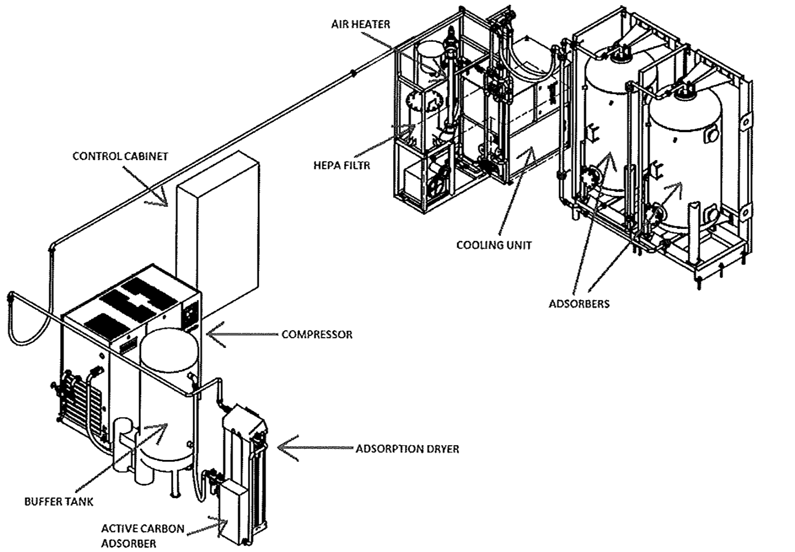} 
   \caption{\emph{Scheme of the distribution of the elements of the RAS at LSC.}}
   \label{fig:RAS}
\end{figure}

The Radon Abatement System at LSC is a system that can provide very low radon-concentration air, of about  $\sim 1$ mBq/m$^{3}$, with a flow from 180 to 220 m$^{3}$/h. For many applications, such level of radioactivity makes this air radon-free (Rn-free) in practice. The daily consumption of both N$_{2}$ and LN$_{2}$ (Liquid nitrogen) is larger than what LSC can produce. The RAS is used to create large Rn-free spaces that are safe for people to work. The Rn-free air is cheaper and also reduces the dependence from external LN$_{2}$ suppliers.

The RAS machine was built by the company Apparatuses Technologies Construction (ATEKO a.s.) from Czech Republic. For this reason, the radon-free air is also named ATEKO-air \cite{ref-journal2}.

Other underground laboratories have similar facilities to reduce the radon activity \cite{ref-journal30}, \cite{ref-proceeding5}.  For example, there are several installations made by ATEKO - Table \ref{tab:atekoInst}  \cite{ref-url2}.

The RAS takes the ambient air and compresses it to 10 bar. Then, the air is dried, filtered and sent to the cooling unit, where the air is cooled to a temperature of -60 \textordmasculine C. In the next step, the air is moved to two adsorption columns with 1 m$^{3}$ volume each, where active carbon captures the radon atoms. Finally, the air temperature is adjusted to be heated back to room temperature and filtered again. For a schematic view of the RAS system, see Figure \ref{fig:RAS}.

 At the LSC, there are two different types of piping for the ATEKO-air: a pressurized Rn-free air line (7 bar) with several outlets at the Underground Laboratory, and an atmospheric pressure line used by the NEXT experiment, with over-pressures from 0 Pa to 5000 Pa (50 mbar). The RAS has been delivering continuously Rn-free air during the last 4 years, with a downtime of 0.2\% due to maintenance.

\begin{table}
  \caption{\emph{ATEKO s.a. installations in Underground Laboratories.}}\label{tab:atekoInst}
\begin{center}    
    \begin{tabular}{llccc}
        \hline
        {\bf Underground Lab.}& {\bf Country} &{\bf Average Rn (Bq/m$^{3}$)}&{\bf Year}&{\bf Air Flow [m$^{3}$/h]}\\
        \hline
        {Modane} & {France} &15& {2004} & {150}\\
        {Gran Sasso} & {Italy} &80& {2011} & {150}\\
        {Gran Sasso} & {Italy} &80& {2012} & {220}\\
        {Yangyang} & {Korea} &40&{2015} & {120}\\
        {LSC} & {Spain} &70& {2015} & {220}\\
        {SURF} & {USA} &310& {2017} & {300}\\
        {Jin Ping} & {China} &40& {2019} & {300}\\
        \hline
    \end{tabular}
 \end{center}
 \end{table}

\section{ELECTROSTATIC RADON MONITOR}
\begin{figure}[htbp] 
   \centering
   \includegraphics[width=7cm]{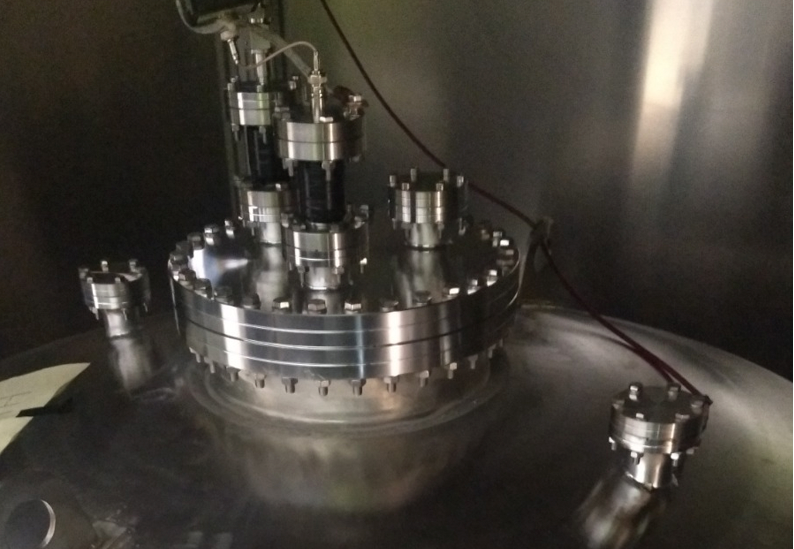}\hspace{10px}
    \includegraphics[width=6cm]{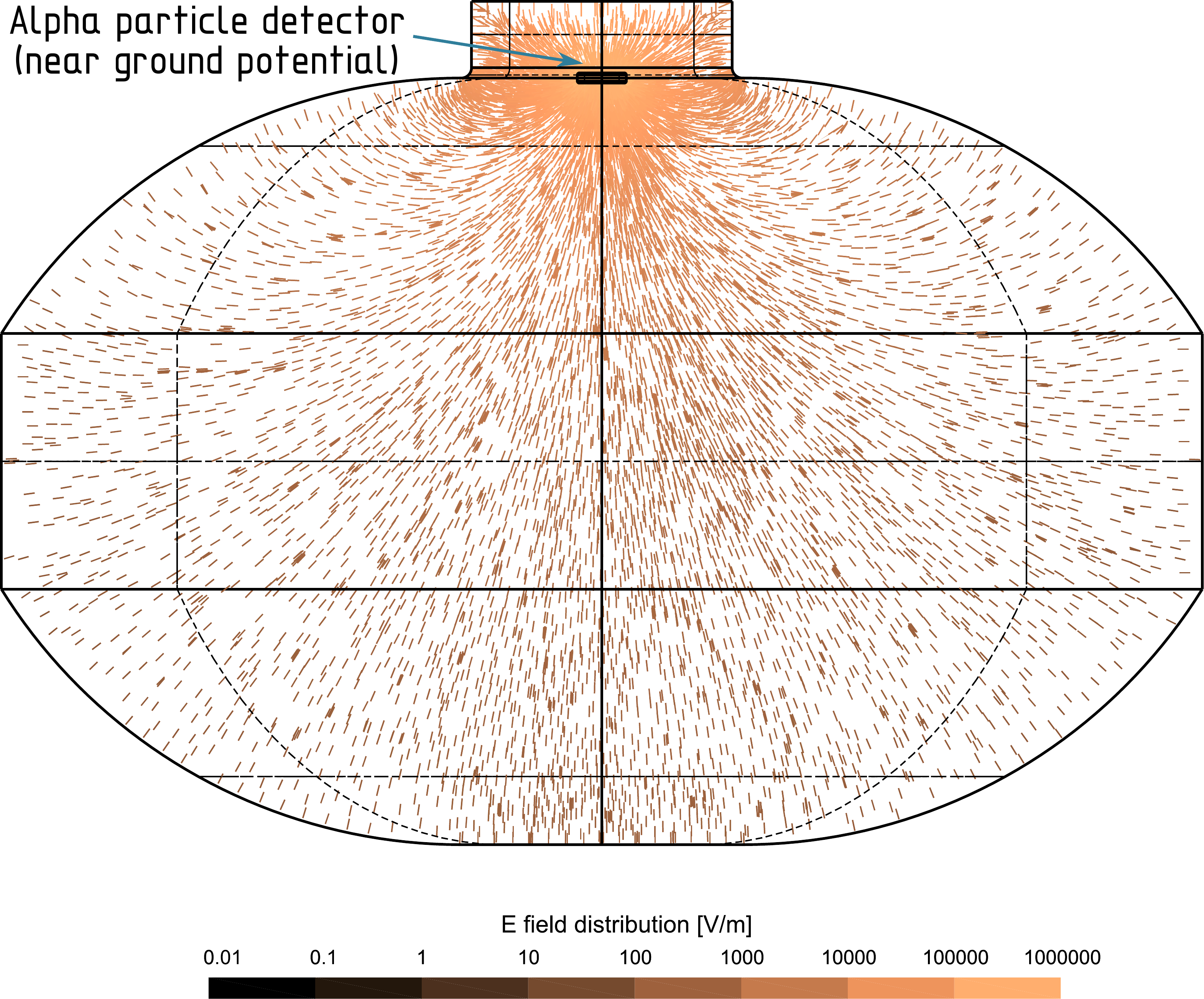} 
   \caption{\emph{Left: Picture of the ERM. Right: Simulation of the electric field lines inside the ERM detector.}}
   \label{fig:ERM}
\end{figure}

The Electrostatic Radon Monitor (ERM) \cite{ref-journal2} is a low background electrostatic detector that can measure in real time activities $<1$ mBq/m$^3$ of $^{222}$Rn in the air using a Silicon $\alpha$-particle Detector from Canberra model PIPS 1200-35-100AM as alpha detector. To evaluate the performance of the ATEKO-air, this air is continuously flowing through the ERM with a slight overpressure of 0.3 barg. Its scheme and a real picture of the top flange is shown in Figure \ref{fig:ERM}. On the right side of the figure, the simulation of the electric field in the detector vessel is shown. The field is created by applying a high voltage (several kV) to the vessel while the alpha detector (black disc in the neck) installed on a top flange is connected to the ground potential. With the simulation, we estimate that $\sim 90\%$ of the field lines cross the detector surface. 

\begin{figure}[htbp] 
   \centering
   \includegraphics[width=13.5cm]{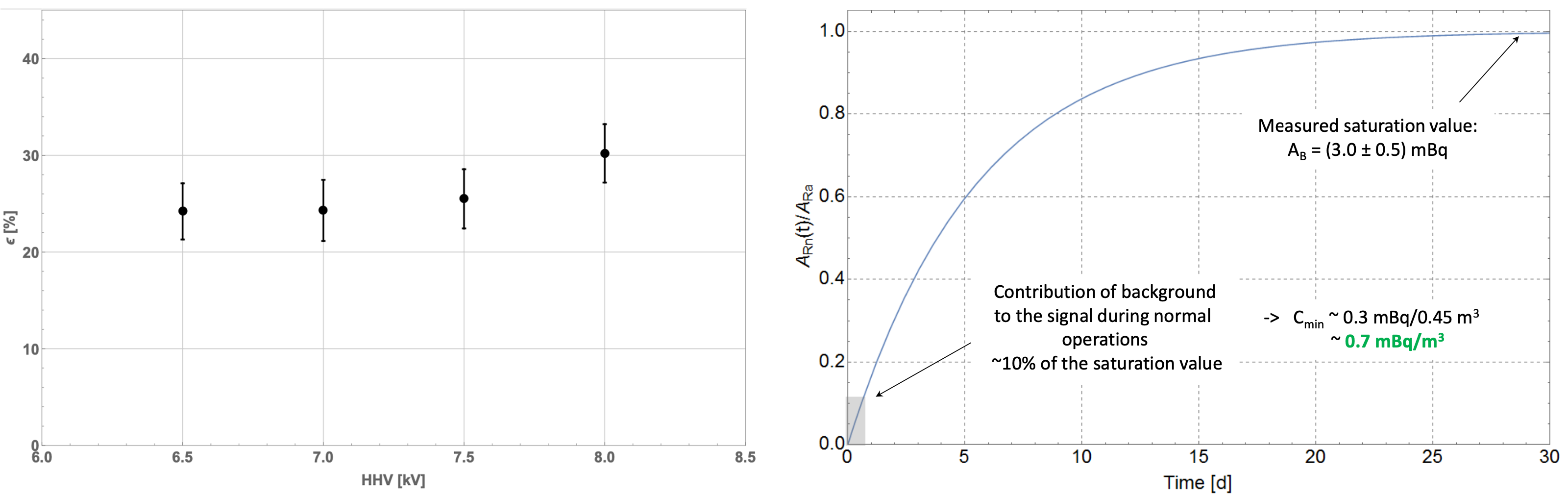}
     \caption{\emph{Left: Overall measured detection efficiency of the ERM for the $^{218}$Po and $^{214}$Po peaks as function of the operational voltage, HHV (High High Voltage). Right: The activity of radon emanated from the ERM detector (in saturation).}}
     \label{fig:ERM_plots}
\end{figure}

The detection efficiency has two contributions: charge collection efficiency and alpha detector efficiency. The charge collection is the process where the $^{218}$Po ions, produced after a $^{222}$Rn decay, are drifted to the alpha detector and collected. The average charge collection efficiency estimated by a detailed simulation of the vessel geometry with a volumetric efficiency of $\sim 90\%$. Some dead volumes (e.g. above the alpha particle detector) contribute to the remaining $10\%$ of the volume (see \cite{ref-journal2} for more details). The intrinsic detection efficiency of the alpha detector is close to  $\sim 50\%$ as the decays occur right at the detector's surface (0 distance). It is actually less than that due to the edge effects (like the detector rim etc.). To measure the detection efficiency, a calibration with a $^{222}$Rn source was performed. The detection efficiency of the ERM is $(30 \pm 4)\%$ for the $^{218}$Po and $^{214}$Po $\alpha$-peaks  at the operational voltage of 8 kV (see Figure \ref{fig:ERM_plots}, left). The detection efficiency is similar in all the peaks of both Polonium isotopes (we tend to refer to peaks - lines - in the energy spectrum having in mind the detection efficiency for particular Po decays). The quoted value refers to $^{214}$Po, which is less sensitive to gas contamination.

To estimate the background of the detector, we closed the flow of ATEKO-air during 1 month to reach the saturation regime - when the measured Rn concentration is actually the activity of emanated Rn (see Figure \ref{fig:ERM_plots}, right). This happens when the detector is sealed (no gas is flushed), and the emanated Rn activity has enough time to build up (saturate) to equilibrium between emanation and decay. The activity of the radon emanated from the detector in saturation is $(3.0 \pm 0.5)$ mBq, which results in a small contribution (about 10 $\%$ of the saturation value) during normal operation. The estimated detection limit of the detector is 0.7 mBq/m$^3$.

\begin{figure}[htbp] 
   \centering
   \includegraphics[width=12cm]{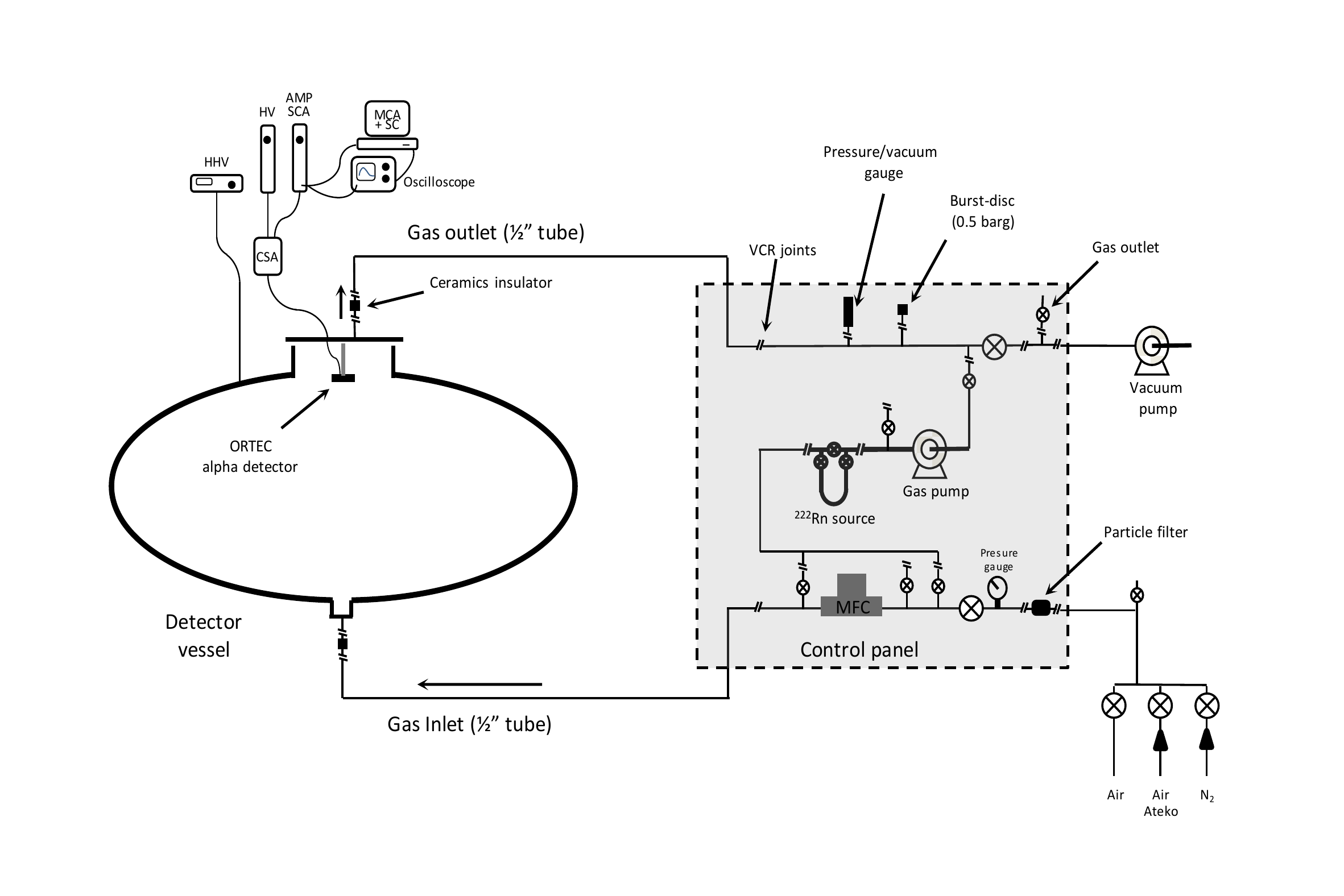}
   \caption{\emph{Schematic of the ERM detector and gas panel.}}
   \label{fig:scheme}
\end{figure}

A gas panel was designed to permit the operation of the ERM detector in various configurations (Figure \ref{fig:scheme} shows the scheme of the detector and the gas panel). There are three different gas inlets: one for N$_{2}$, one for the ATEKO-air directly from the RAS and the other one for different uses. Currently, the detector is configured to input the ATEKO-air that comes back from NEXT after flushing inside the lead castle, around the detector. A built-in $^{226}$Ra calibration source (that emits $^{222}$Rn) is connected to the gas panel. A gas pump is used to force circulation of the air through the detector.

\begin{figure}[htbp] 
   \centering
   \includegraphics[width=14cm]{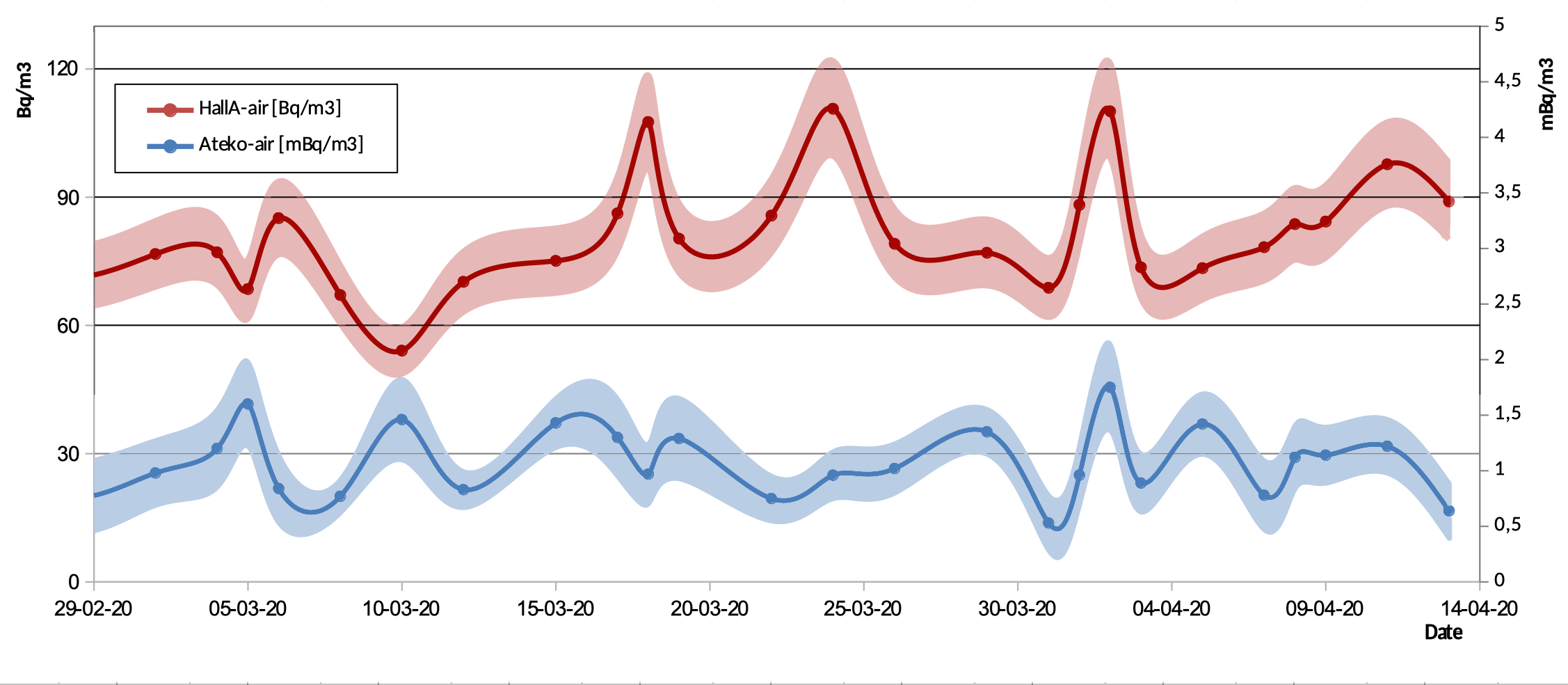} 
   \caption{\emph{Comparison of radon activities vs time. We show the radon activity in Hall-A measured with an AlphaGuard (in red) in units shown in the left side y-axis and with the ERM (in blue) in units shown in the right side y-axis.}}
   \label{fig:Rn_comp}
\end{figure}

To see if there is a dependency of the residual Rn in the ATEKO air on the Rn concentration outside the detector (input air), a study was made (In Figure \ref{fig:Rn_comp}) by comparing the daily activity measured with the AlphaGuards (normal air) and the daily activity measured with the ERM (ATEKO-air). The AlphaGuard detector performs a measurement every 10 minutes, daily mean values were considered. In the case of ERM, some measurements are several days long; in this case we considered the corresponding mean value. The colored areas around the lines include the 1-sigma error. No clear correlation was observed between the Rn content in the ATEKO line and in the laboratory air. For this reason, it is  mandatory to have dedicated detectors - ERM and AlphaGuard, to monitor the activity of the ATEKO-air and the external background, respectively.

\begin{figure}[htbp] 
   \centering
   \includegraphics[width=11cm]{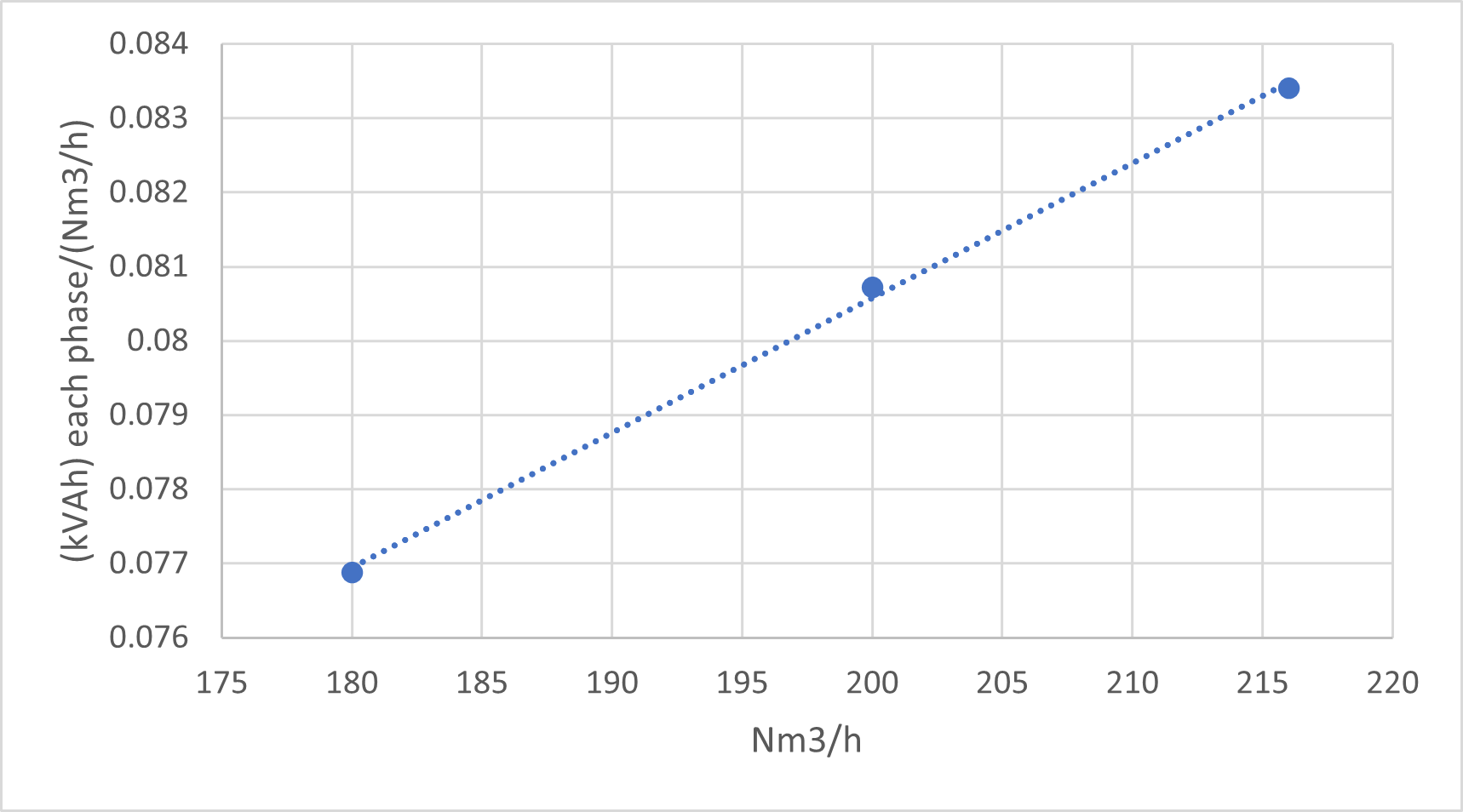} 
   \caption{\emph{Energy consumption required to generate a unit of flux (about 1 Nm3/h) of ATEKO-air depends on the outlet flux level produced by the system}}
   \label{fig:ateko_w}
\end{figure}

The ATEKO-air is very stable in time, demonstrated by calculating the mean value of the activities for the first half of 2020 and 2021. The mean activity was $1.1 \pm 0.2$ mBq/m$^3$, in 2020 and also in 2021. It shows that the performance of the active carbon is good for long time usage. For this measurements the outlet flow from the ATEKO system was 185 m$^3$/h. We also evaluated how much Rn activity varies with the air flow selected in the ATEKO machine. From June 8$^{th}$ to June 15$^{th}$ 2020, the flow of the ATEKO-air was increased to maximum capacity, that is 220 m$^3$/h, with a mean activity of $1.0 \pm 0.2$ mBq/m$^3$ statistically compatible with the usual mean activity. This measurement was long enough to demonstrate no  significant differences in activity between flows. The energy consumption required to generate a unit of flux (about 1 Nm$^3$/h) of ATEKO-air depends on the outlet flux level produced by the system (see Figure \ref{fig:ateko_w}). Therefore, activity is not an important factor to select  the flow of Rn-free air.

\section{ATEKO-AIR TO NEXT}\label{sec:4}

The first experiment  benefiting from the ATEKO-air in the underground lab was NEXT (Neutrino Experiment with a Xenon TPC) experiment \cite{ref-journal5}. NEXT collaboration is trying to identify the nature of neutrino, whether it is a Majorana fermion or not, with the observation of $0\nu\beta\beta$ (Neutrinoless Double-Beta Decay) of $^{136}$Xe, with Q$_{\beta\beta}$ = 2.5 MeV. The detector is a time projection chamber filled with xenon gas at pressure (7-15 bar), and two separated detection planes: the energy plane - 
which is used to measure the energy of events with PhotoMultiplier Tubes (PMT) with energy resolution (ER) $<1\%$, and the tracking plane - which is used to track the electrons in the gas with Silicon photomultipliers (SiPM). The main sources of background to this experiment are the decays of two isotopes: $^{214}$Bi (2.447 MeV, intensity: 1.57\%) and $^{208}$Tl (2.614 MeV, intensity: 100\%) -  as shown in Figure \ref{fig:rad_chains}, with gammas very close to Q$_{\beta\beta}$.

An external Lead Castle (with 200 mm thickness) was built to shield the detector from the radioactive gamma background of the laboratory \cite{ref-journal6}. At the same time, a system to flush with ATEKO-air was developed. Inside the Castle a diffusor with 8 air outputs (each one with 8 horizontal holes and 1 vertical) was mounted to flush the detector with ATEKO-air (see below).

\subsection{Firsts Tests of the Castle}
In 2014, before the RAS and the installation of NEXT-White detector and the diffusor, the initial plan was to reduce the Radon background inside the Castle by building it as air-tight as possible and flowing a small volume of N$_{2}$ inside. To evaluate this design, several tests (from 29/10/2014 to 12/02/2015) \cite{ref-proceeding1} were done with a high pressure N$_{2}$ bottle, changing the flow and purging between measurements. These tests were made with the Castle completely built, with all the joints in place but before placing the vessel of NEXT-White on the pedestal. Therefore, the empty holes for the pipes of the Castle were covered with High Density Polyethylene (PE-500).

\begin{figure}[h] 
   \centering
   \includegraphics[width=13cm]{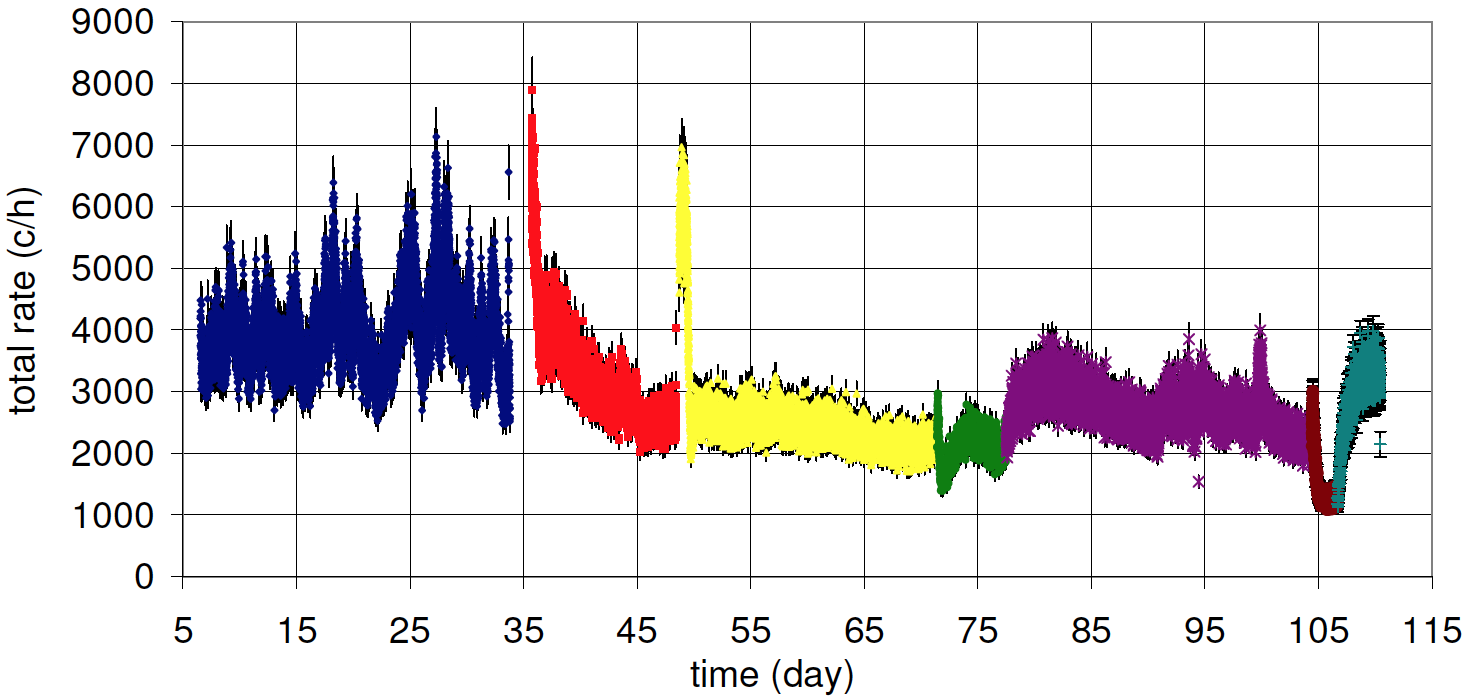} 
   \caption{\emph{NaI(Tl) total rate as a function of time for different tests: {\bf blue} - Castle closed; {\bf red} - Castle closed with improved tightness; {\bf yellow} - Castle closed with 7h N$_{2}$ purge at 1800 l/h; {\bf green} - Castle closed with N$_{2}$ purge + constant N$_{2}$ flux 180 l/h; {\bf violet} - Castle closed without N$_{2}$ flux; {\bf brown} - Castle closed with N$_{2}$ purge + N$_{2}$ flux 900 l/h; {\bf turquoise} - Castle closed without N$_{2}$ flux.}}
   \label{fig:2012-CastleTest}
\end{figure}

Two different detectors were placed inside the Castle for these measurements: one was an AlphaGuard Saphimo P2000 and the other was a 3'' $\times$ 3'' NaI(Tl) detector. The NaI(Tl) instrumented crystal is a low background detector that was connected to a Canberra 2005 preamplifier, a Tennelec linear amplifier and a Canberra 8701 ADC, and finally read by an Arduino board. The rate measured with the NaI(Tl) detector is shown in Figure \ref{fig:2012-CastleTest}. During these days, the Rn activity in Hall A was in the range of 60 $-$ 80 Bq/m$^{3}$.

To evaluate the contribution of the Castle, different setups were prepared:
\begin{enumerate}
\item \emph{Castle open}. This measurement was taken before closing the Castle to have a reference value of the event rate. With the Castle open, the space between both sides of the Castle is 2.9 m, therefore it was only partially shielded from the gamma background coming from the laboratory. This measurement is not shown in Figure \ref{fig:2012-CastleTest}.
\item \emph{Castle closed}, drawn in blue in Figure \ref{fig:2012-CastleTest}. Measurement done after closing the Castle.
\item \emph{Better closed Castle}, drawn in red. Measurement done after improving the air-tightness with additional joints, blocking the holes, etc.
\item \emph{After N$_{2}$ purge}, drawn in yellow. A purge was done with N$_{2}$ with a flow of 1800 l/h for 7 hours to remove the Rn inside the Castle. 
\item \emph{N$_{2}$ purge + flux 180 l/h}, drawn in green. A second purge was made but, after this, a constant flux of 180 l/h was injected into the Castle.
\item \emph{Without N$_{2}$ flux}, drawn in violet. Measurement done after stopping the constant flow of N$_{2}$ in setup 5.
\item \emph{N$_{2}$ purge + flux 900 l/h}, drawn in brown. Same conditions as in setup 5 but with a larger flux of 900 l/h.
\item \emph{Without N$_{2}$ flux}, drawn in turquoise. Same as in setup 6, measurement made after stopping the N$_{2}$ flux.
\end{enumerate}
The results of these measurements are presented in Figure \ref{fig:2012-CastleTest} and Table \ref{tab:CastleRn}. In the Figure, NaI(Tl) total rate as function of time for different tests (in different colors) are presented. In the third column of the table the average rate for each test is calculated. In the fourth column the Radon activity measured in short time windows for each test is shown.

\begin{table}
\begin{center}    
\small
\caption{ {\emph{Measurements of the contribution of the NEXT Castle in different tests.}}}
    \begin{tabular}{cccc}
        \hline
        { } & {\bf Setup} & {\bf Rate with NaI (Hz)} & {\bf Activity with AlphaGuard (Bq/m$^{3}$)$^\star$} \\
        \hline
        {1} & {Castle open} & {$59.56 \pm 0.01$} & { $79 \pm 28$ }\\
       
        {2} & {Closed Castle} & {$1.089 \pm 0.001$} & {$79 \pm 28$ }\\
      
        {3} & {Better closed Castle} & {$0.694  \pm 0.002$} & {$30 \pm 11$ }\\
        
        {4} & {After N$_{2}$ purge} & {$0.658 \pm 0.001$} & {$30 \pm 12$ }\\
        
        {5} & {N$_{2}$ purge + flux 180 l/h} & {$0.600 \pm 0.001$} & {$22 \pm 9$ }\\
        
        {6} & {Without N$_{2}$ flux} & {$0.638 \pm 0.001$} & {$26 \pm 10$ }\\
        
        {7} & {N$_{2}$ purge + flux 900 l/h} & {$0.350 \pm 0.002$} & {$4 \pm 3$ }\\
 
        {8} & {Without N$_{2}$ flux} & {$0.894 \pm 0.002$} & {$47 \pm 15$ }\\
        \hline
    \end{tabular}
    \label{tab:CastleRn}
 \end{center}
 \begin{center}    
  $^\star$\tiny{The AlphaGuard measurements were very short, only to obtain the order of magnitude.}
  \end{center}
 \end{table}

In setup 7, we observe a large background reduction. But for this measurement the AlphaGuard detector is at its limit of sensitivity and the NaI(Tl) could be contaminated by gammas from other sources. Therefore, the use of Radon-free gas and more precise detectors are needed to improve the measurement.

\subsection{Connection of NEXT to RAS}

fter the installation of the RAS and the ERM, the possibility of a cheaper and Rn-free air produced at LSC was the best option. Rn-free air is produced at LSC and can be precisely measured with ERM. Additionally, more flexible safety protocols around the Castle can be followed because there is no risk of suffocation. The connection between the RAS and NEXT can be divided into three different sections: Part I - LSC pipes system, Part II - the NEXT pipes outside the Castle, and Part III -  the NEXT pipes inside the Castle.

\subsubsection{Description of Part I - the LSC pipe system}
NEXT experiment and the RAS are in opposite parts of the Hall A, separated by about 60 meters. Therefore, a pipe crossing the Hall A is used to deliver the air. The installation is made of AISI-316 Stainless Steel pipe with d$ = 4''$ (101,6 mm), welded, minimizing the number of joints. The length of this pipe is 59 m. All joints are using flat flanges, according to norm EN-1092-1 Type 1 PN-2.5. These flanges use rubber elastomeric gaskets. For the air distribution, two manual ball valves are mounted at the outlet of the RAS. In parallel to this installation, there is a pipe to return the air from NEXT Castle to the ERM with d$ = 1''$ (25,4 mm) and made of AISI-316 Stainless Steel, completely welded. This pipe has a length of 59 meters and has unions with threaded fittings, according to norm DIN-2999.

\subsubsection{Description of  Part II - the NEXT experiment pipe system}
\begin{figure}[htbp] 
   \centering
   \includegraphics[width=10cm]{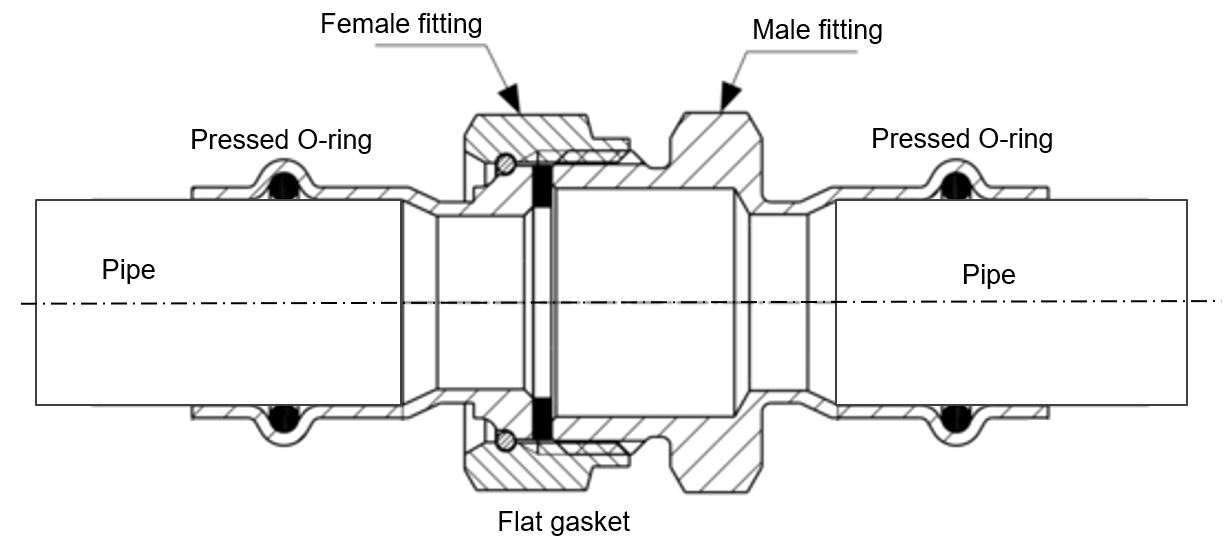} 
   \caption{\emph{Lateral section of the pressed joints and threaded joints, pressfitting INOXPRES}}
   \label{fig:Union-NEXT}
\end{figure}

The system consists of a pipe with diameter d$ = $1'' 1/2 (38.1 mm) of AISI-316 Stainless Steel and several flexible pipes with d = 40 or 50 mm of PVC (SPRINGVIN IBERFLEX). Pressed joints and threaded joints (type Pressfitting, INOXPRESS) are used to connect these pipes (see Figure \ref{fig:Union-NEXT}). All these joints are using elastomeric gaskets of EPDM. For air distribution, several ball valves are employed. We used flexible pipes to allow the opening of the external Lead Castle and best suited to seismic vibrations affecting the installation.

The NEXT pipe system includes several gauges:
\begin{enumerate}
\item A flow meter Produal IVL10N with two output signals (for ventilation and air conditioning). The signals correspond to air flow velocity and temperature inside the air duct.
\item A differential pressure sensor WIKA Model A2G-50 (for ventilation and air conditioning). This sensor measures the air pressure at three different points of the installation using solenoid valves: 
\subitem a) the pressure in the Rn-free air inlet line, 
\subitem b) the pressure in the pipe entering in the NEXT Castle and,
\subitem c) the pressure inside the Castle, around the detector.
\item A manual flow-control valve to control the air flow to the Castle.
\item Several manual ball valves located in different 'bypasses', installed for tests purposes for the RAS and the ERM.
\item The system is monitored by an Arduino ARDBOX 20, from Industrial Shields.
\end{enumerate}

\subsubsection{Description of Part III - the diffusor, the detector and the Castle}
\begin{figure}[htbp] 
   \centering
   \includegraphics[width=10cm]{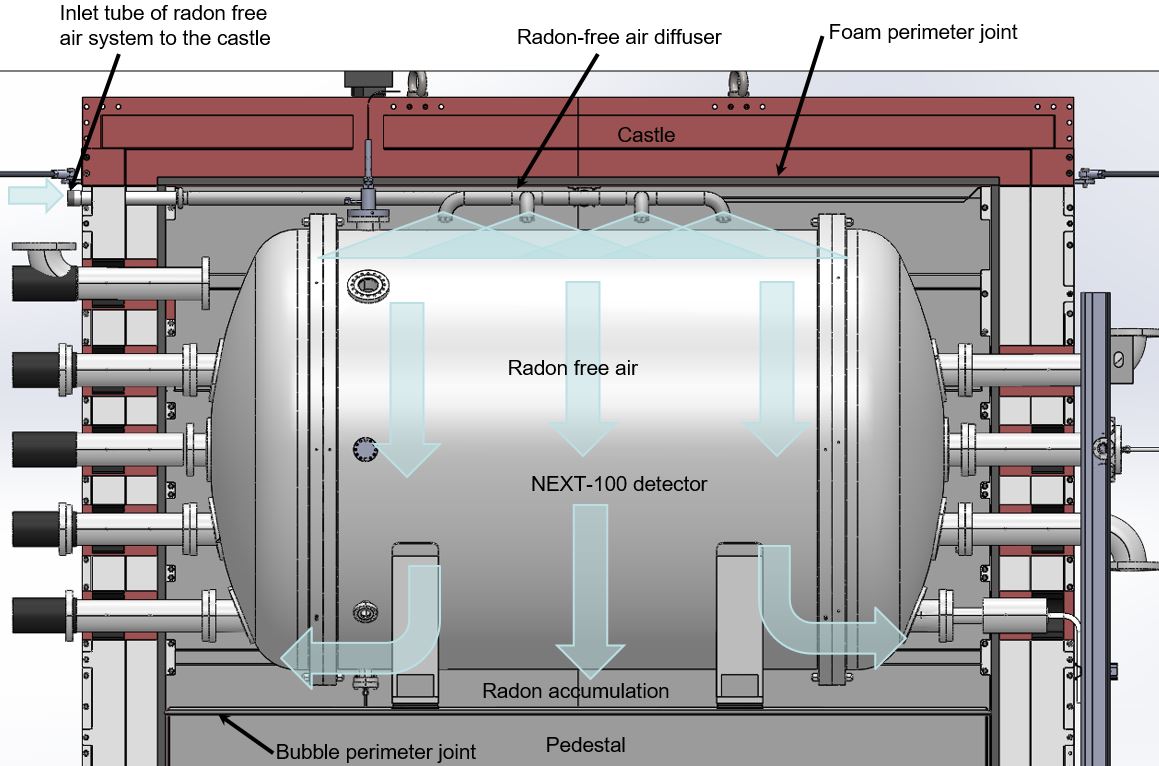} 
   \caption{\emph{Side view of the Rn-free air diffusor, the Castle and the NEXT100 detector.}}
   \label{fig:Sideview-NEXT}
\end{figure}

The air diffusor is placed inside the Castle, above the NEXT detector. This diffusor is designed to homogeneously distribute the ATEKO-air inside the Castle and displace the normal air (see Figure \ref{fig:Sideview-NEXT}). In this Figure, the air flow injected by the diffusor is shown in light blue. This diffusor is designed to be used for both NEXT-White and NEXT100 detectors.

\begin{figure}[htbp] 
   \centering
   \includegraphics[width=10cm]{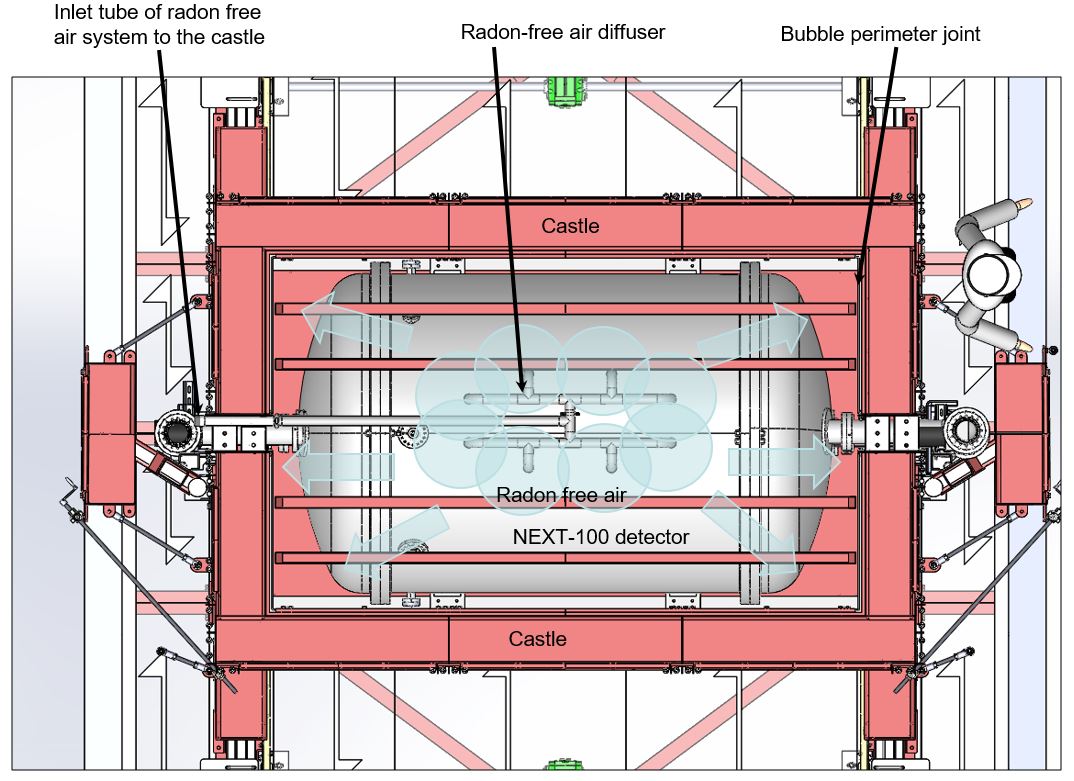} 
   \caption{\emph{Top view of the closed Castle with a scheme of the distribution of the Rn-free air.}}
   \label{fig:Topview-NEXT}
\end{figure}

We show an upper cut (without the roof of the Castle) of the diffusor and the NEXT-100 vessel in Figure \ref{fig:Topview-NEXT}. The flows of injected air from the 8 tubes of the diffusor were marked with blue circles, above the detector. The Castle (in red) is divided into two mobile parts. Both sides are independent and can be moved to get access to the detector. For this purpose, a system with rails, wagon wheels, and two electric motors to open and close the Castle was implemented.

\begin{figure}[htbp] 
   \centering
   \includegraphics[width=10cm]{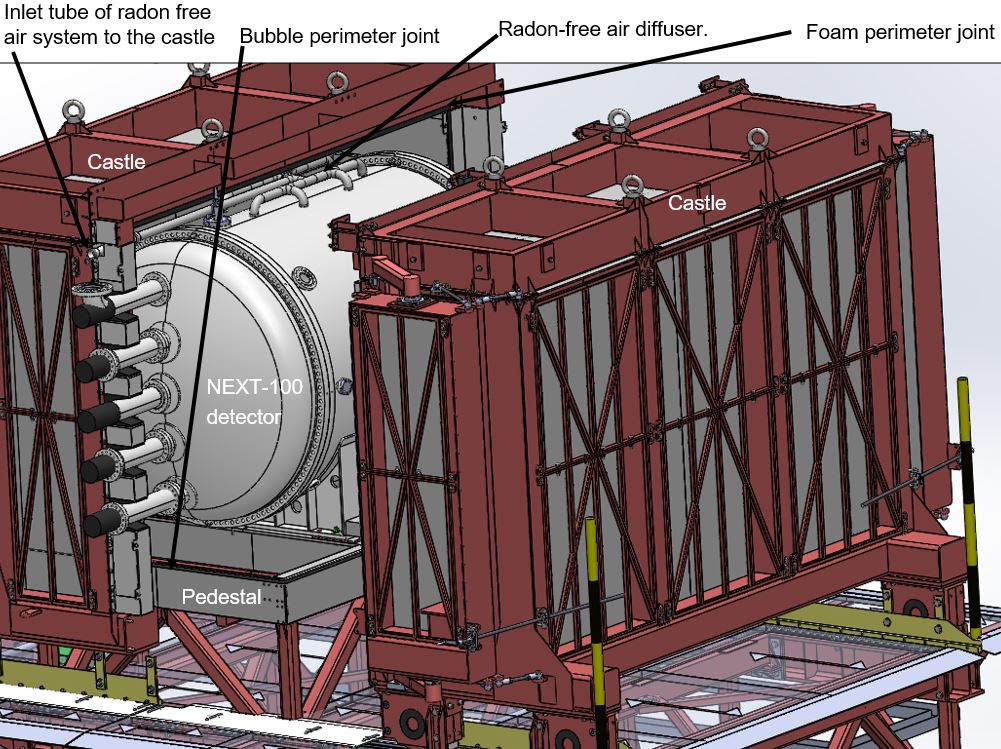} 
   \caption{\emph{NEXT detector. Details of the connection system with the ATEKO machine.}}
   \label{fig:Castle-NEXT}
\end{figure}

The Castle is shown with the right side opened in Figure \ref{fig:Castle-NEXT} , where five of the ten pipes coming from the RAS can be seen. The ten pipes (DN-100) are crossing the two mobile parts of the lead of the Castle. To improve the Castle airtightness and reduce the gamma background around the detector, special lead bricks were machined to perfectly seal the space around the pipes. With these bricks in place, the Castle can be opened and closed without manually removing any other part of the shielding.  There are foam fillings in all the joints between the two mobile sides of the Castle and there is a bubble joint in the union between the pedestal and the two mobile parts. All these joints are necessary to keep the Castle air-tight.

\begin{figure}[htbp] 
   \centering
   \includegraphics[width=7cm]{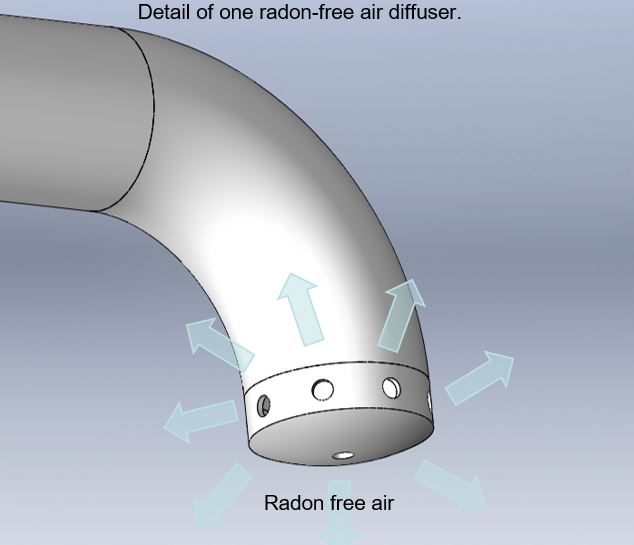} 
   \caption{\emph{Diffusor detail: 8 perimetral holes and 1 central, all of them of diameter d$ = $5 mm.}}
   \label{fig:Diffusor-NEXT}
\end{figure}

A closer view on a diffusor is shown in Figure \ref{fig:Diffusor-NEXT}. It has 8 perimetral holes and 1 central, all of them of diameter d$ = $5 mm. Radon tends to accumulate above the pedestal. For this reason, these diffusor tubes are designed to displace it using a large ATEKO-air flow injection. The air will escape from the Castle through the external perimeter of the pedestal and also through the small holes between the pipes and the lead (small leaks).

\subsection{Tests of the pipes and the Castle}
\begin{figure}[htbp] 
   \centering
   \includegraphics[width=13cm]{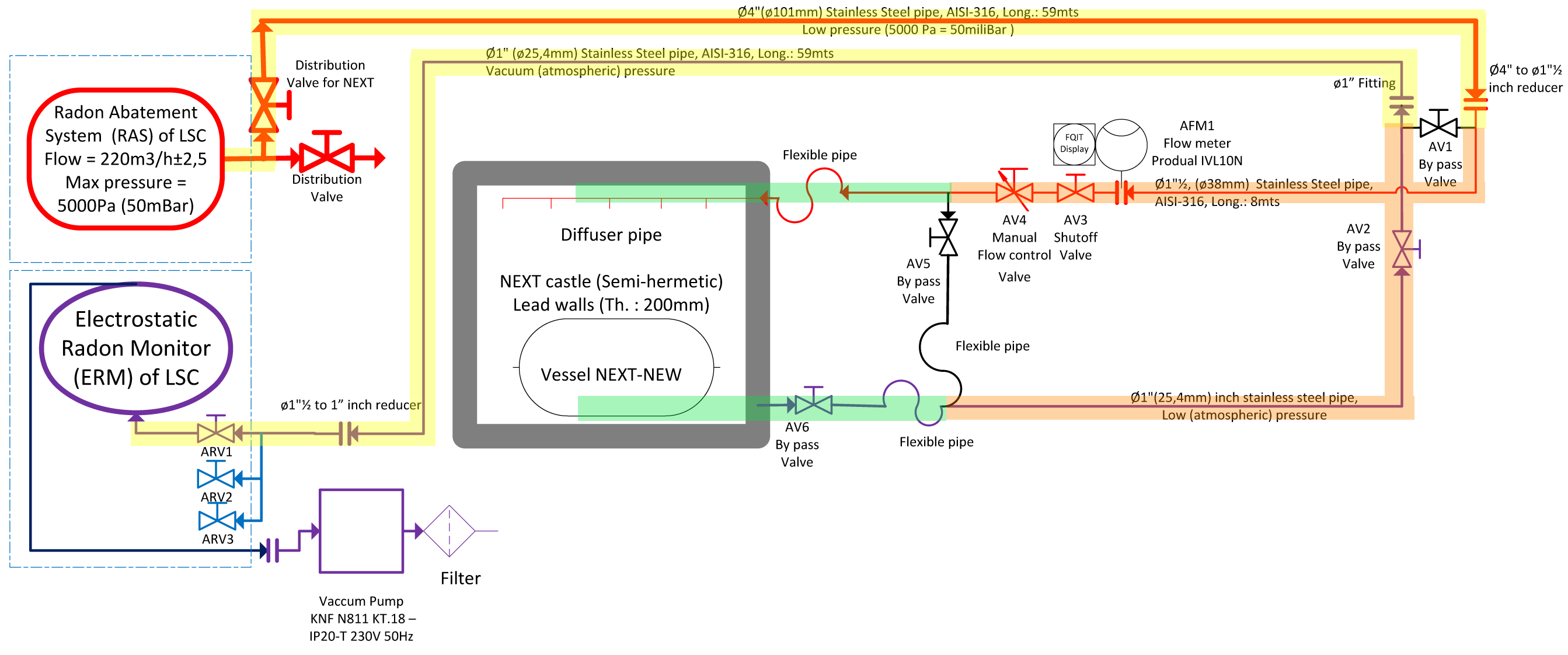} 
   \caption{\emph{Paths used for the three different measurements of Rn activity}}
   \label{fig:Test-Paths}
\end{figure}

Due to the large distance between the RAS and the ERM, long pipes were needed: pipes from the LSC setup are 59 m long and the ones from setup NEXT are 8 meters long. Considering also the return line, the total distance is 134 m. To evaluate as precisely as possible the radon contribution of the different parts of the system, 3 different tests were performed (see Figure \ref{fig:Test-Paths} and Table \ref{tab:next}). It is important to evaluate the contribution of all parts because some pressed tubes were used instead of welded stainless steel pipes. Pressed tubes are cheaper and easier to mount.

{\bf Test-1} (Part I - path marked in yellow in Figure \ref{fig:Test-Paths}) takes the ATEKO-air and crosses all the Hall A and returns to the ERM to measure the Rn activity of the air. To isolate this circuit, AV1 bypass valve was used. The total length of this circuit is 118 m.

{\bf Test-2} (Part I and Part II - paths marked in yellow and orange in Figure \ref{fig:Test-Paths}) includes all  pipes but not entering inside the Castle. In this case, AV1 was closed and AV5 opened. The total length of this circuit is 132 m.

{\bf Test-3} (Part I, Part II and Part III - paths marked in yellow, orange and green in Figure \ref{fig:Test-Paths}) uses the final setup used in NEXT. The ATEKO-air enters in the Castle and, through the diffusors, baths NEXT-White detector. To complete the circuit, a small open pipe at the base of the pedestal takes the air into the return line. The open pipe is placed where the Rn accumulates with the air flowing (lower part of the Castle). With this setup, there is no overpressure so we use a pump to force the circulation of the air towards the ERM. AV1 and AV5 are closed. The total length of this circuit is 134 m.

\begin{table}
\caption{\label{tab:next} \emph{Activity Rn measured with ERM for the Tests.}}
\begin{center}    
    \begin{tabular}{cccc}
        \hline
        {\bf Test-1} & {\bf Test-2} & {\bf Test-3 - I} & {\bf Test-3 - II} \\
        \hline
        {$\leq 1.3$ mBq/m$^{3}$} & {$\leq 4.3$ mBq/m$^{3}$} & {45.0 mBq/m$^{3}$} & {31.0 mBq/m$^{3}$ }\\
        \hline
    \end{tabular}
 \end{center}
 \end{table}

Two levels of flow at the entrance of the Castle were used in {\bf Test-3}. {\bf Test-3-I} (conditions similar as in Test-1 and Test-2), was performed with a flow of $\sim$ 400 l/m. For {\bf Test-3-II}, a larger flow of $\sim$ 550 l/min was (see Figure \ref{fig:Test-Paths}) used. The Rn-free air significantly reduces the amount of radon (factor larger than 1000) in the neighborhood of the NEXT-100 vessel, a number that varies with the air flow. 

\begin{figure}[htbp] 
   \centering
   \includegraphics[width=12cm]{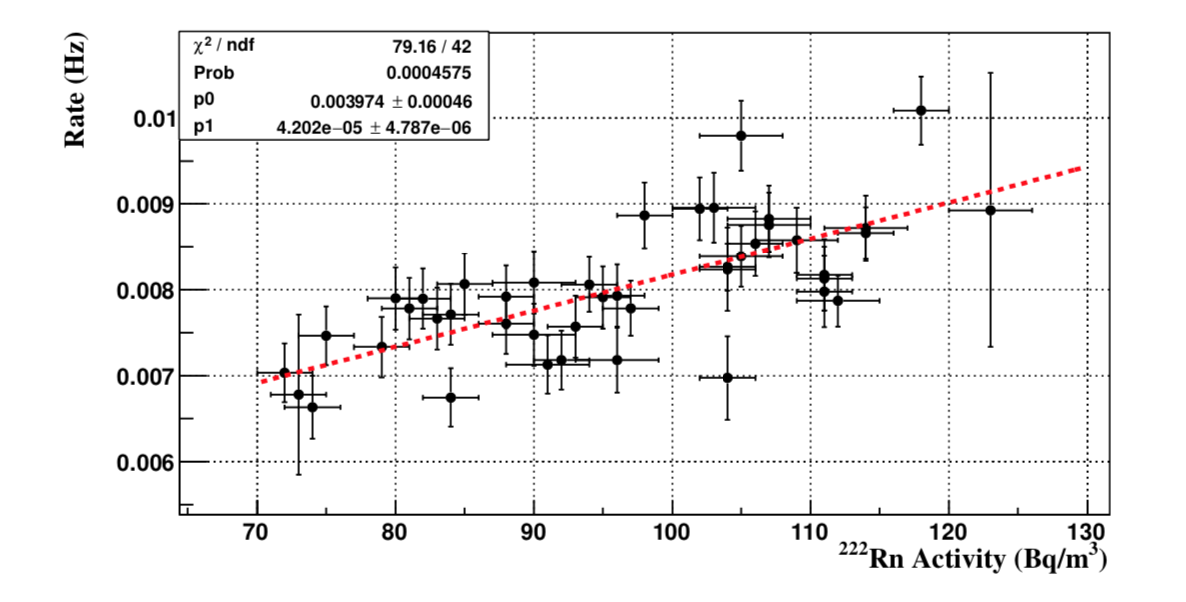} 
   \caption{\emph{Daily Radon activity vs Background rate in NEXT-White \cite{ref-journal7}.}}
   \label{fig:NEXT_Bkg_Rn1}
\end{figure}

\begin{figure}[htbp] 
   \centering
   \includegraphics[width=11cm]{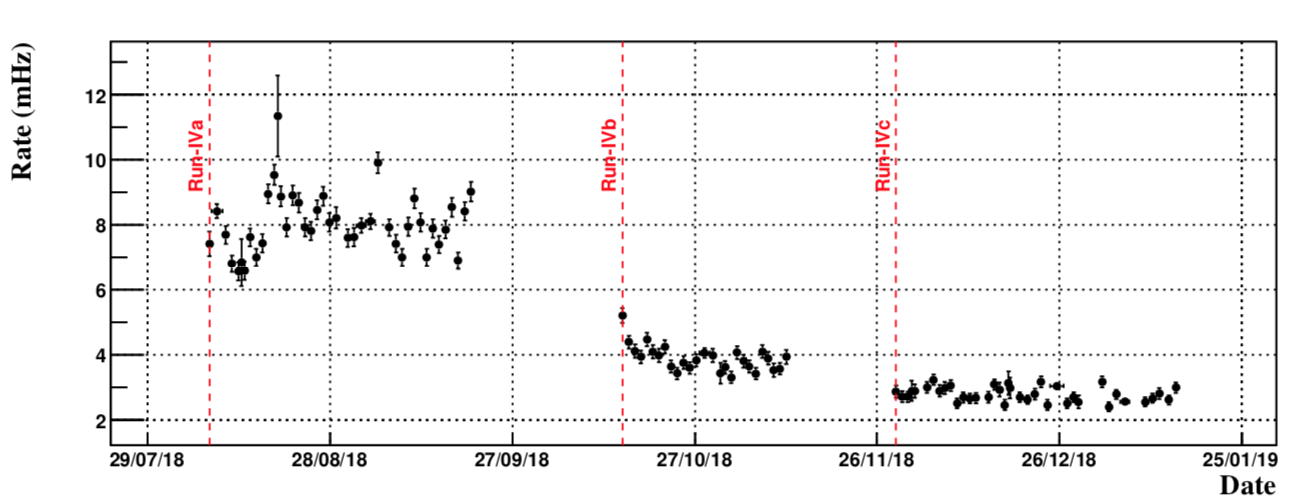} 
   \caption{\emph{Evolution of the total background in NEXT-White detector. The vertical dashed red lines indicate the beginning of each run  \cite{ref-journal7}.}}
   \label{fig:NEXT_Bkg_Rn2}
\end{figure}

Before the ATEKO-air diffusor installation was completed, a study of the influence of the Rn concentration outside the detector (between 60 and 100 Bq/m$^{3}$ - measured with an AlphaGuard detector) on NEXT background was performed with the NEXT-White detector (see Figure \ref{fig:NEXT_Bkg_Rn1}) \cite{ref-journal7}. A linear correlation may be used to estimate the background rate in NEXT detector and the Rn concentration outside the vessel. Extrapolating the red dashed line to a Rn-free scenario, the expected background rate was $3.97\pm 0.46$ mHz. During the background run of NEXT-White, data has been taken first without ATEKO-air (run IVa) and later with ATEKO-air flowing around the detector (run IVb). Without ATEKO-air, the background rate was $8.00\pm 0.05$ (stat) $\pm 0.07$ (sys) mHz, and with ATEKO-air was $3.90\pm 0.05$ (stat) $\pm 0.04$ (sys) mHz (see Figure \ref{fig:NEXT_Bkg_Rn2}) \cite{ref-journal7}. The result of run IVb and the previous extrapolation are in a very good agreement. In Run IVc an additionally second inner lead shield was mounted. In this run the background rate was reduced to $2.78\pm 0.03$ (stat) $\pm 0.03$ (sys) mHz.


\section{ATEKO-AIR TO ULBS}\label{sec:5}

\begin{figure}[htb]
\centering
   \begin{minipage}{0.5\textwidth}
     \centering
     \includegraphics[width=1.\linewidth]{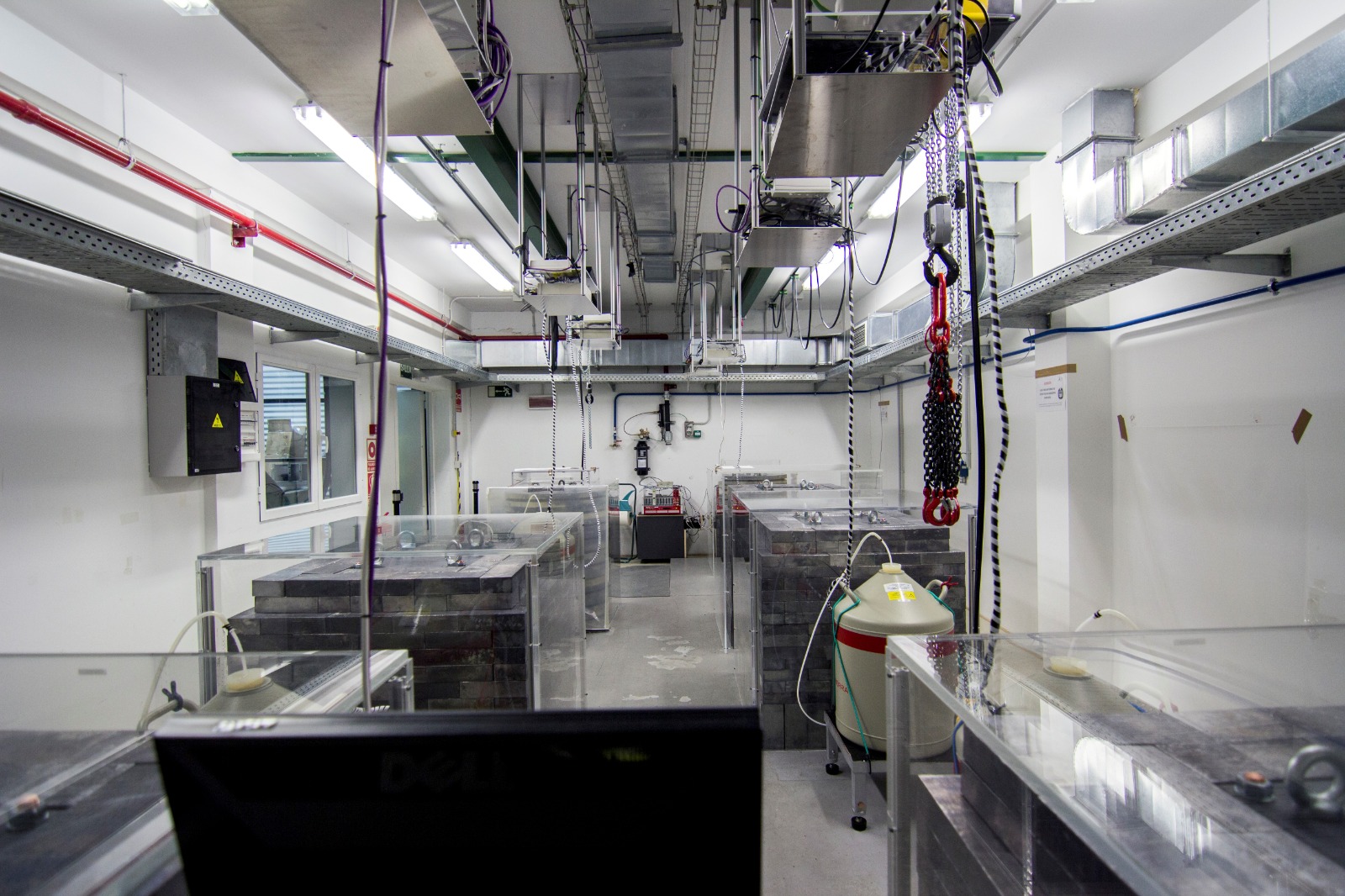}
     \caption{\emph{ULBS laboratory in Hall C of LSC.}}\label{fig:ulbs}
     \vspace{7mm}
     \includegraphics[width=1.\linewidth]{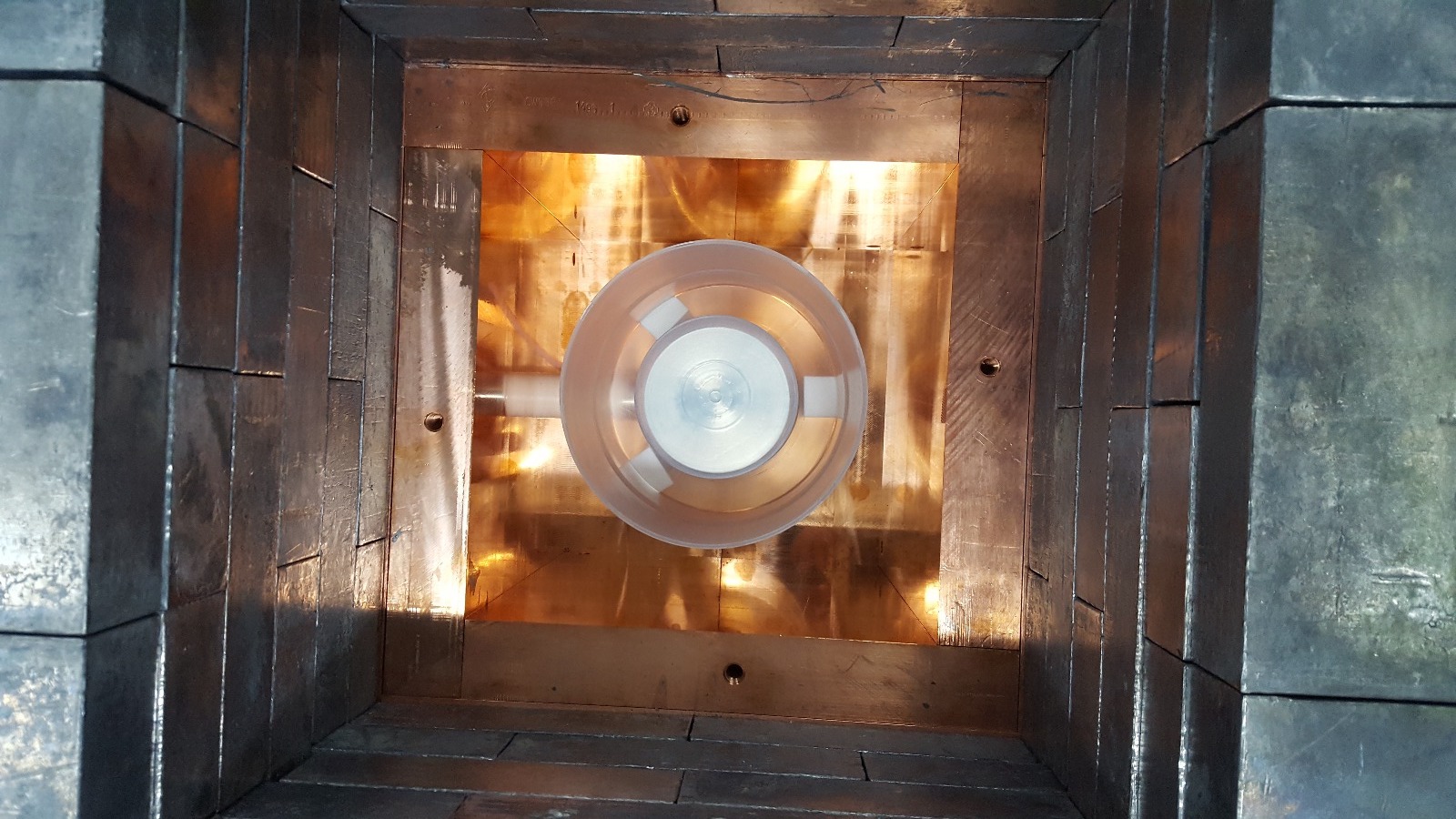}
     \caption{\emph{Sample volume inside the Cu shielding.}}\label{fig:inhpge}
   \end{minipage}
\end{figure}

The Ultra-Low Background Service (ULBS), working since 2010 in Hall C of LSC (Figure \ref{fig:ulbs}), is offering a high-quality screening facility to experiments \cite{ref-journal4}. At present, it is equipped with six 2 kg p-type coaxial High Purity Germanium (HPGe) detectors with $\sim 100\%$ relative efficiency, shielded with 20 cm of lead with a low contamination in $^{210}$Pb ($<30$ mBq/kg). An internal 10 cm of OFHC (Oxygen-Free High thermal Conductivity) copper layer completes the shielding. The ULBS carries out mainly radio-purity assay for experiments running at LSC and collaborations \cite{ref-journal21}, \cite{ref-proceeding2}, \cite{ref-journal22}, \cite{ref-proceeding4} and also contributes to SK-Gd \cite{ref-journal10} and HK \cite{ref-journal11} experiments in Kamioka. 

\begin{figure}[htbp] 
   \centering
   \includegraphics[width=12cm]{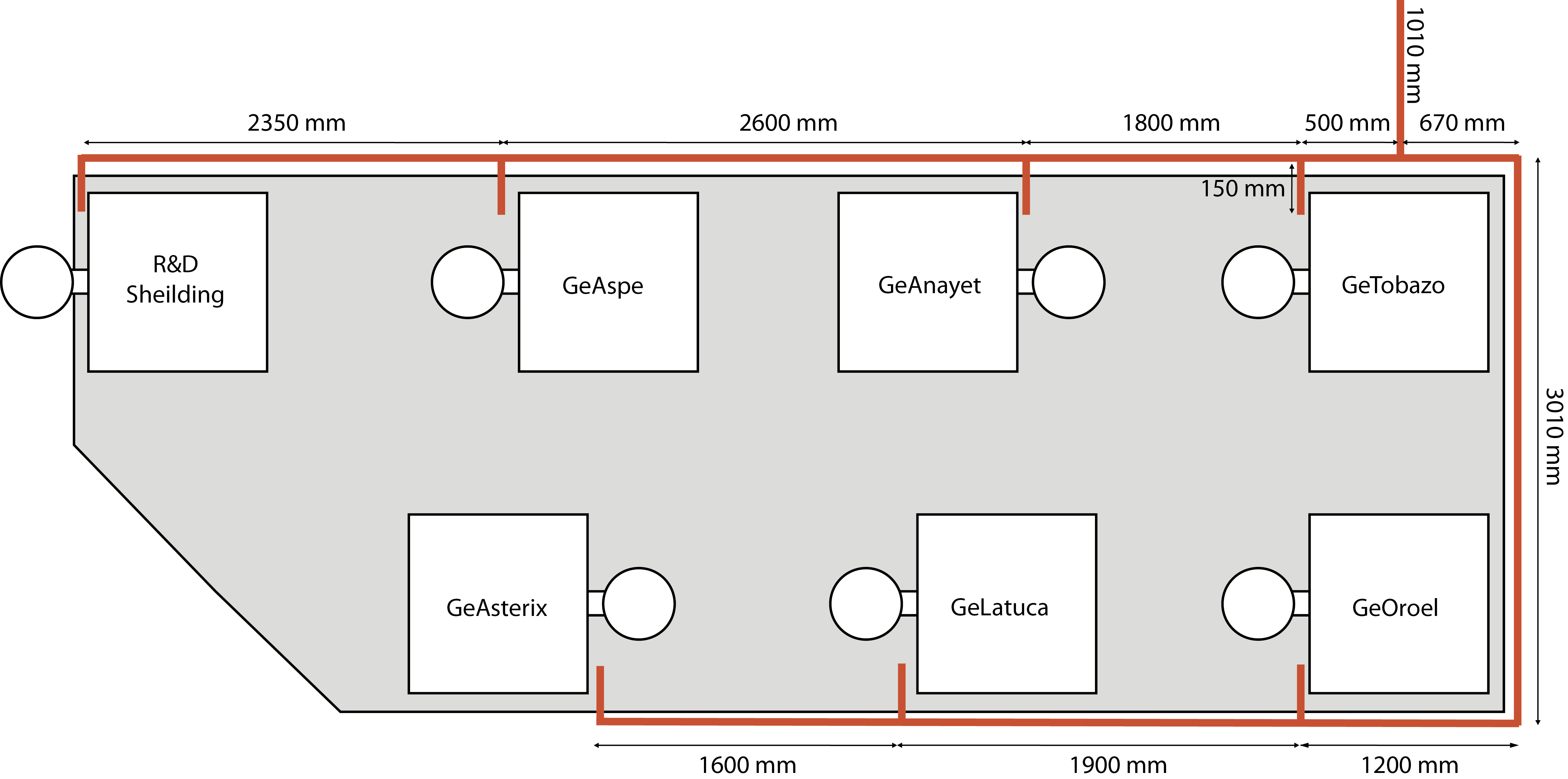} 
   \caption{\emph{Schematic of HPGe positioning in Hall C. In red, the installation to supply ATEKO-Air to each detector.}}
   \label{fig:inst}
\end{figure}

\begin{figure}[htbp] 
   \centering
    \includegraphics[width=0.2\linewidth, angle=0,valign=t]{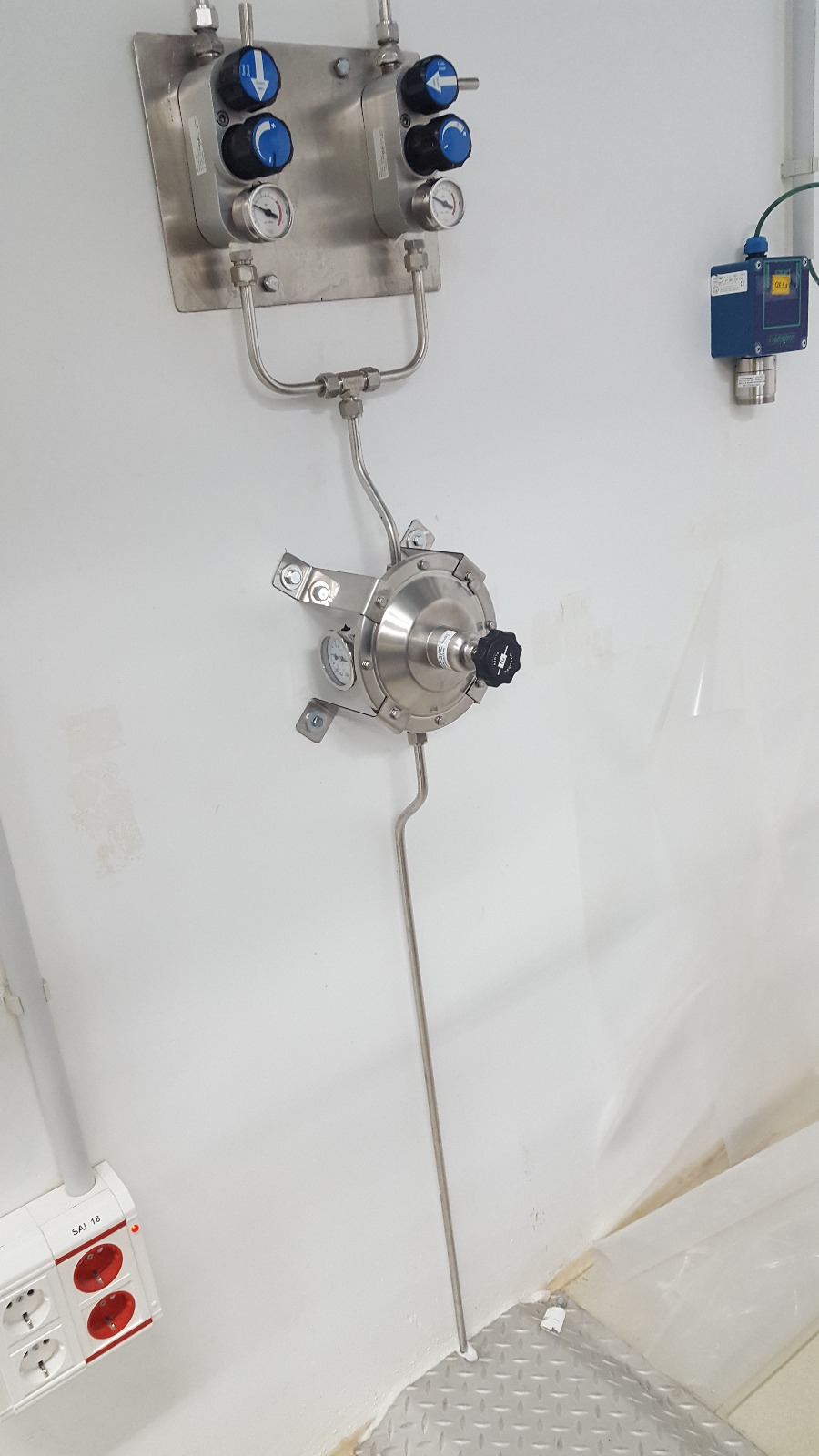}
    \caption{\emph{ATEKO-Air main line in\\ Hall C.}}\label{fig:hpgairm}
\end{figure}

\begin{figure}[htbp] 
   \centering
     \includegraphics[width=0.2\linewidth, angle=0,valign=t]{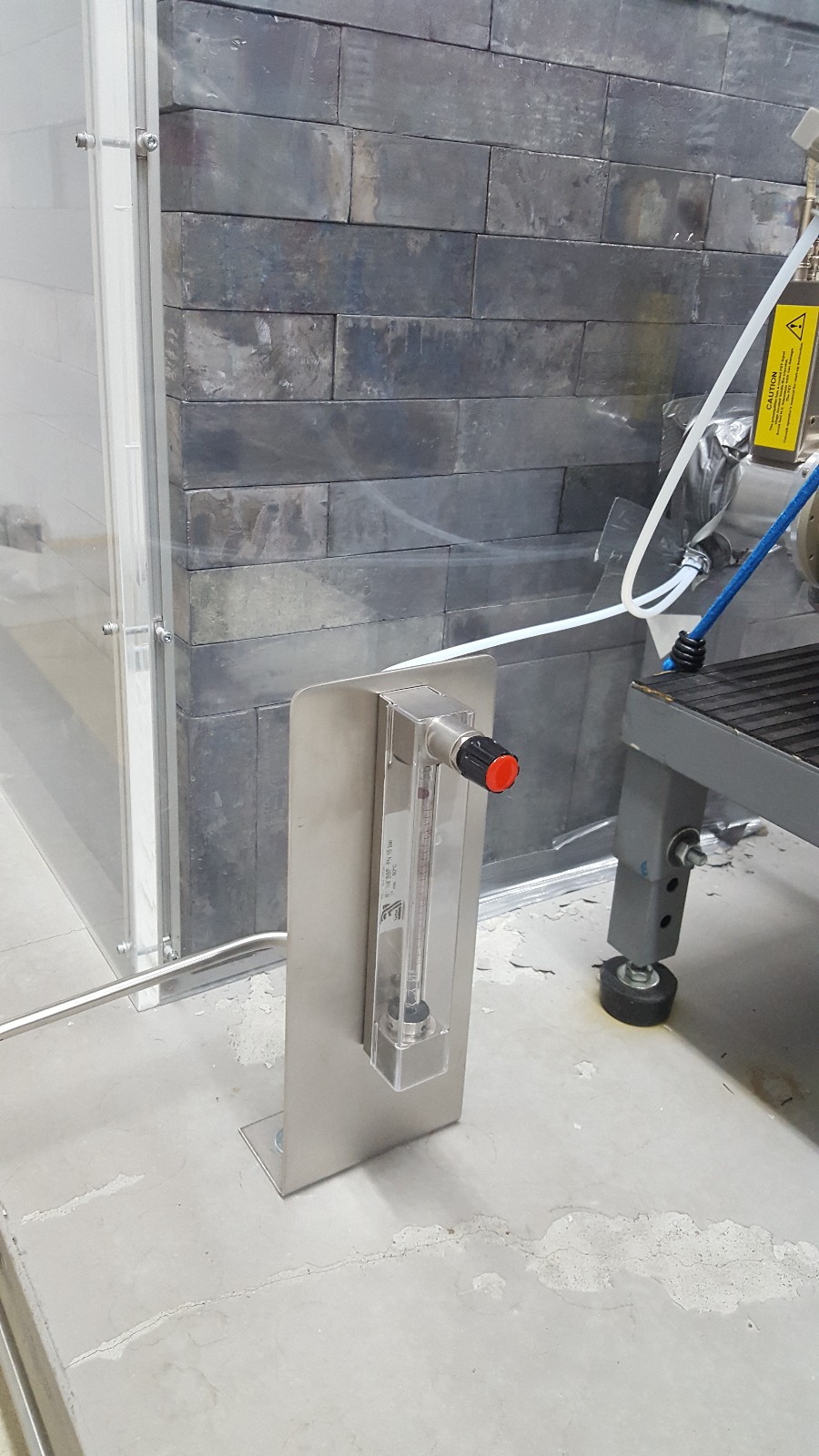}
     \caption{\emph{ATEKO-Air Rotameter with valve\\ to measure and regulate air flow rate.}}\label{fig:hpgeair}
\end{figure}

Long measurements (about 30 days) are necessary to reach sensitivity levels of 1 mBq/kg or less, with the condition that the Radon contained in the air inside the HPGe-detector shielding is at the level of 1 mBq/m$^{3}$ or less. For this purpose, each detector has an extra layer of shielding made of methacrylate that externally surrounds the lead shielding. To impede the exterior air (and the Rn) from entering inside the shielding, a slight overpressure is created inside the copper shielding. The inner volume available for sample measurement (the inside of the Cu shielding) has the geometry of a cube with side 350 mm (see Figure \ref{fig:inhpge}). This gives a volume of $\sim 43$ l.

Two systems are used to reduce the Rn concentration inside the HPGe-detector copper shielding. The first system uses the nitrogen that is evaporating from the detector's dewar. Each HPGe-detector in LSC is equipped with a 30 l dewar for LN$_{2}$ necessary for the detector operation. LN$_{2}$ has to be refilled each week. Using a 6 mm diameter Teflon tube, we flush continuously $\sim 54$ l/h inside the detector the N$_{2}$ gas which is evaporating from the dewar. This value was obtained by measuring (with a graded stick) the height of LN$_{2}$ and calculate the evaporated gas considering that each liter of LN$_{2}$ will produce  $\sim 696$ liters of gaseous N$_{2}$. Before the RAS, this setup was the only anti-Radon system.

\begin{table}
  \caption{\emph{Background counts in the 2 setups: 2016-setup with a flux of N$_{2}$ from dewar evaporation of 54L/h; 2021-setup with a flux of 274 l/h made by combined N$_{2}$ from dewar evaporation and ATEKO-Air.}}\label{tab:comp}
\begin{center}    
    \begin{tabular}{ccccc}
        \hline
        {}& {\bf Flux}& {\bf Gas}&{\bf Int. rate [40 - 2700 keV]} & {\bf Int. rate [609.32 keV peak]} \\
         {}& {\bf [l/h]}& &{\bf [cts/kg/day]} & {\bf [cts/kg/day]} \\
        \hline
        {2016 setup}& 54 & N$_{2}$ &165.3 & {2.9}\\
        {2021 setup}&274 & N$_{2}$+Air&  141.8 & 0.4\\
        \hline
    \end{tabular}
 \end{center}
 \end{table}

The second system, using the RAS, reduces even more the total background of the HPGe detectors. A fixed installation was specially built in 2019 to flow ATEKO-air inside the shielding of the detectors (Figure \ref{fig:inst}). We did some tests and we concluded that there is no difference in the detector's background rate when using the LN$_{2}$ evaporation or ATEKO-air if the flow inside the shielding has the same value. Currently, the flow is 220 l/h in each detector. This flow rate can be regulated with a valve for each detector and the flux is measured with a rotameter (see Figures \ref{fig:hpgairm} and \ref{fig:hpgeair}).

To optimize the use of the ATEKO-air and obtain a proper Radon reduction we are using both systems. The total flux (ATEKO-air + N$_{2}$) flushed inside the detector is $\sim 274$ l/h so the air is changed more than 6 times per hour in the sample volume.

\begin{figure}[!htb] 
   \centering
   \includegraphics[width=10.cm]{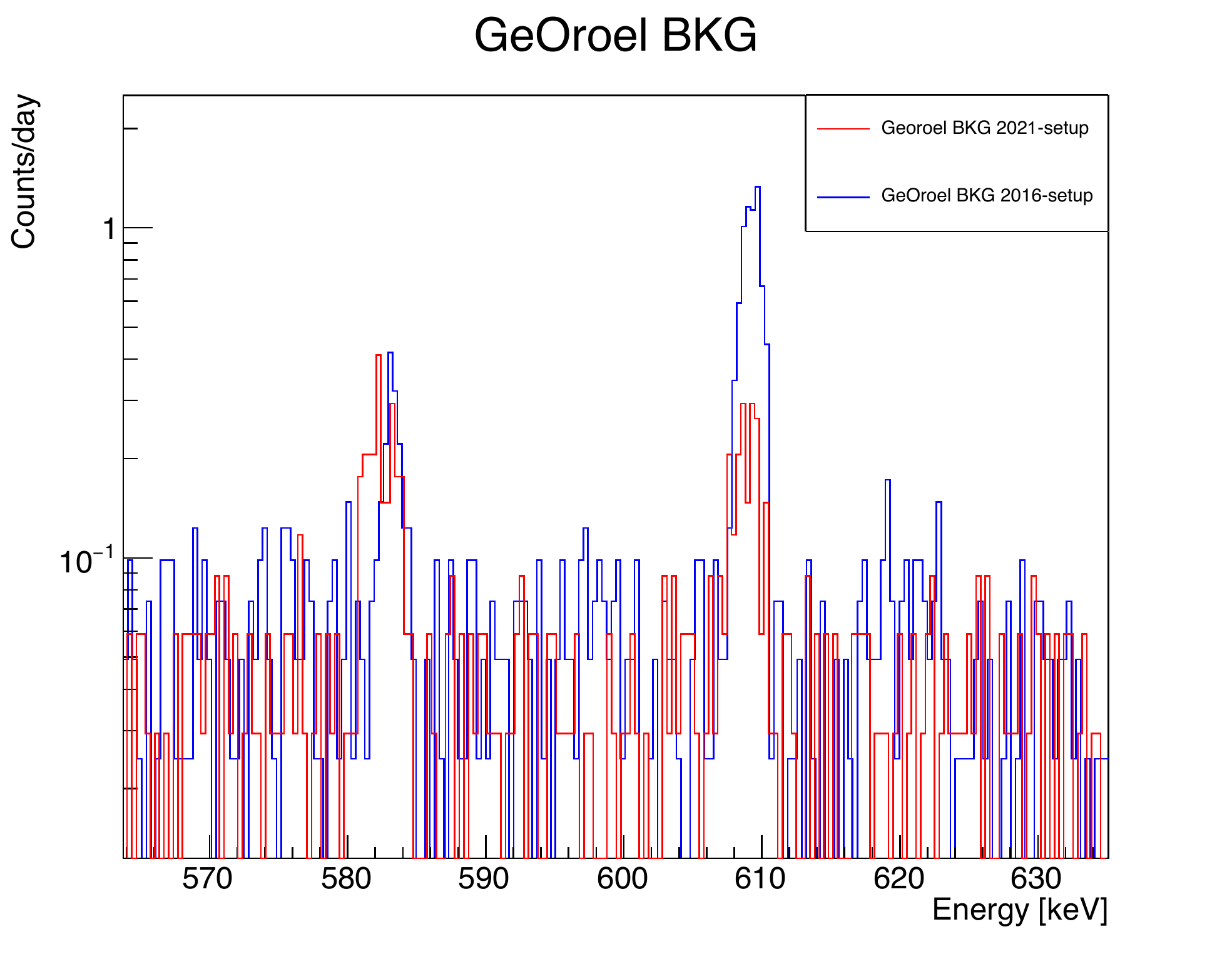} 
   \caption{\emph{GeOroel background comparison,  $^{214}$Bi 609.32 keV peak of reduction. In blue 2016 setup vs. in red 2021-setup.}}
   \label{fig:georoel}
\end{figure}

To show the background reduction, we compare (Table \ref{tab:comp} and Figure \ref{fig:georoel}) the rates of GeOroel (one of the best HPGe’s of LSC) with the old (2016) and the current setup, i.e., N$_{2}$ evaporated from the dewar versus ATEKO-air injected and N$_{2}$ evaporated from the dewar.

\section{ATEKO-AIR TO CROSS EXPERIMENT}\label{sec:6}

The goal of the Cryogenic Rare-event Observatory with Surface Sensitivity (CROSS) \cite{ref-journal8}, \cite{ref-journal9}, \cite{ref-journal12} is the development of a technology capable of investigating lepton number violation and the nature of neutrino with unprecedented sensitivity, by searching for neutrinoless double beta decay $0\nu\beta\beta$ of two promising isotopes ($^{100}$Mo and $^{130}$Te) with the bolometric approach. The choices of the candidates and of the compounds for CROSS are based on years of development of the bolometric search for  $0\nu\beta\beta$ decay projects: MIBETA, Cuoricino, CUORE, LUCIFER, CUPID-0, LUMINEU and CUPID-Mo (see \cite{ref-journal8} and bibliography inside). The baseline solution is background reduction by the scintillating-bolometer technology, as proposed in the CUPID project \cite{ref-journal91}, \cite{ref-journal92}. CROSS can provide an adequate background abatement with a major simplification of the detector structure and of the readout. 

A dilution refrigerator with an experimental volume of about $\sim 150$ l was installed in April 2019 in the LSC in the framework of the CROSS project. The CROSS key idea is to provide the bolometric detection technique, ideally tailored to the study of this rare nuclear transition as it features high energy resolution, large efficiency and wide flexibility in the detector material choice –- with an additional decisive characteristic: an effective pulse-shape-discrimination (PSD) capability, enabling the rejection of events from surface radioactive impurities and other background-inducing phenomena. CROSS is planning to install a first demonstrator of the CROSS technology in this cryogenic set-up \cite{ref-journal8}. It will consist of 32 Li$_{2}$MoO$_{4}$ crystals grown with molybdenum enriched in $^{100}$Mo at $> 95\%$ level with CROSS technology \cite{ref-journal9}. The region of interest (ROI) is around 3034 keV. Each crystal is a cube with 45 mm side, for a mass of 0.28 kg. This new detector will pave the way to bolometric experiments with low background levels (see Table \ref{tab:demonstrator} \cite{ref-journal8}) to make possible future large searches penetrating in prospects the normal hierarchy of the neutrino masses. 

 \begin{figure}[htb] 
   \centering
    \includegraphics[width=5.35cm]{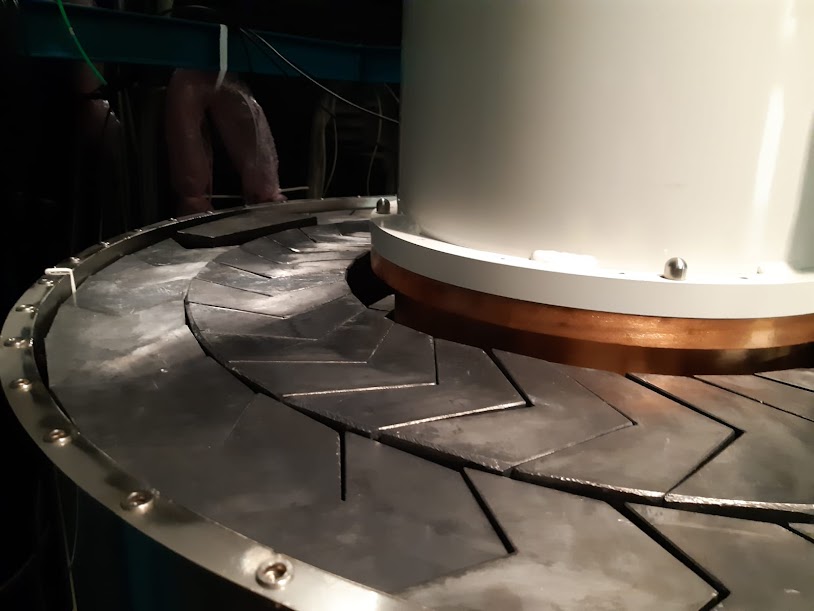}\hspace{4px}
   \includegraphics[width=3cm]{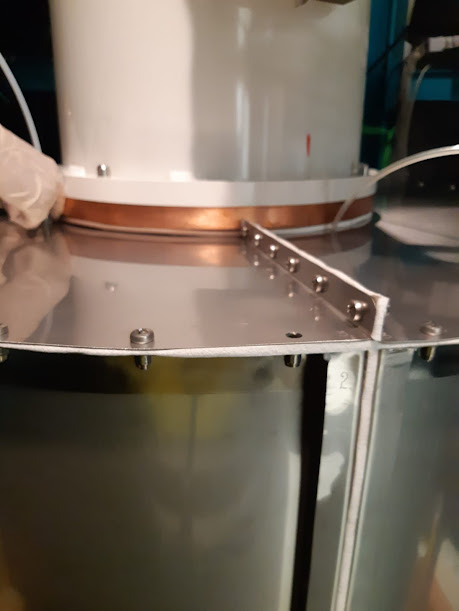} \hspace{2px}
    \includegraphics[width=3cm]{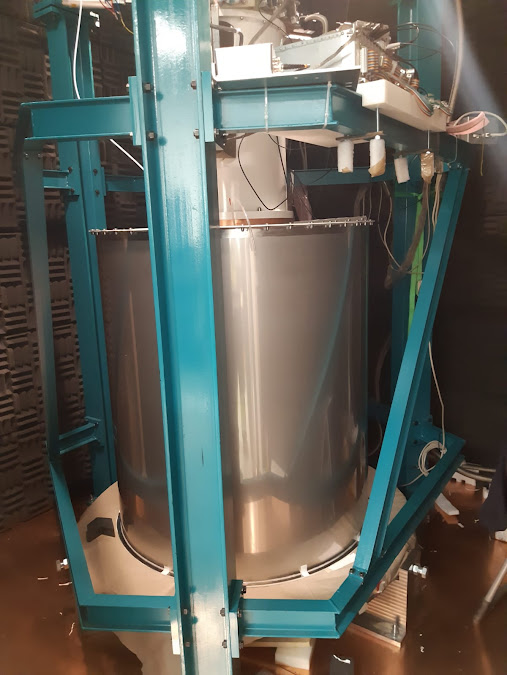}
   \caption{\emph{CROSS Anti-Radon Shield.}}
   \label{fig:RnShield}
\end{figure}

 \begin{table}
 \footnotesize
 \caption{\label{tab:demonstrator} \emph{Sensitivity of the CROSS demonstrator to be operated in the cryogenic facility of the Canfranc underground laboratory, under two different hypotheses of background level. The efficiency is 75\%, compatible with the crystal size of each element. The interval of limits on the effective Majorana neutrino mass $m_{\beta \beta}$ accounts for a recent compilation of nuclear-matrix-element calculations \cite{ref-journal8}.}}
\centering
\begin{tabular}{cccccc}
\hline
{\bf Background} & {\bf Live} & {\bf Background} & {\bf Feldman-Cousins} & {\bf Half-life} & {\bf m$_{\beta\beta}$} \\
{\bf level} & {\bf time} & {\bf counts} & {\bf count limit} & {\bf limit [y]}& {\bf limits [meV]} \\
{\bf [counts/(keV kg y)]} & {\bf [y]} & {\bf in ROI}  & {\bf (90\% C.L.)} &{\bf  (90\% C.L.)} & {\bf (90\% C.L.)}  \\
\hline
$10^{-2}$ & 2 & 1.40 & 3.6 & $8.5\times10^{24}$ & 124--222 \\
$10^{-3}$ & 2 & 0.14 & 2.5 & $1.2\times10^{25}$ & 103--185 \\
$10^{-2}$ & 5 & 3.60 & 4.6 & $1.7\times10^{25}$ & 88--159 \\
$10^{-3}$ & 5 & 0.36 & 2.7 & $2.8\times10^{25}$ & 68--122 \\
\hline
\end{tabular}
\end{table}

To reach such low levels of background all external contributions must be mitigated. One of these background sources \cite{ref-journal8} is the Rn in the air outside the experiment ($69.0\pm 0.3$) Bq/m$^{3}$ as we shown above. To remove this background, the system is similar to the one used in the HPGe detectors: an extra tight shielding made of stainless steel was constructed to surround the detector's cylindrical lead shielding. ATEKO-air is inserted inside this last layer to create a slight overpressure and impede the external air that contains Radon to enter and produce background. 

The shielding design is a cylinder made of a 1 mm Stainless Steel sheet with internal diameter D = 990 mm and 1200 mm hight (see Figure \ref{fig:RnShield}). The shield is made of several components that can be easily assembled. Silicone gaskets and PTFE are used to maintain the ensemble tight. As the free volume that remains inside this anti-Radon shield is relatively small ($\sim 45$ l), a flux of 280 l/h of ATEKO-air is sufficient to remove the Rn background to the order of 1 mBq/m$^{3}$. The ATEKO-air installation is very similar to that installed in ULBS (Hall C), a Rotameter with a valve will be used to regulate the Rn-free air flux inside the detector. The shield was mounted in the fall 2021 and will be tested in the following CROSS run starting in December 2021.

\section{CURRENT PLANS}\label{sec:7}

The Rn-free air line became a much needed service at the underground lab and will be further exploited in the near future, with new science and technology projects.

NEXT-100, an upgraded version of NEXT-White, consists of a larger vessel TPC filled with 100 kg of $^{136}$Xe-enriched Xe at 15 bar. Some improvements to the NEXT Castle and the corresponding ATEKO-air input system are in progress: remove the painting to reduce radioactive background and improve air sealing. Also, because a larger TPC will be mounted (lower volume of air inside the Castle) we can expect an important reduction of the radioactive background.

The Radon-free air supply is also essential for the Copper Electroforming Services (CES) \cite{ref-proceeding3} recently installed in the LSC Clean Room. A major program of improving the service was done recently to obtain the highest radiopurity of the copper produced in Canfranc, currently tested with components for DAMIC experiment \cite{ref-url1}, and for additive printing.  In the near future, it is planned to test the Radon-free air as an inert gas inside the electrolytic bath instead of nitrogen gas. 

The Rn-free air from RAS is necessary in the LSC microbiology and molecular biology laboratory. Currently, five experiments have been approved to start their works underground. These experiments are focused on studies of phenotypic and molecular adaptations of biological models, such as microorganisms or small multicellular eukaryotic organisms in low radiation background environment. The RAS will be used to minimize the Rn environmental background, as done in the physics experiments, to fully monitor the background conditions in the experiments.

\section{CONCLUSIONS }\label{sec:8}

The combination of Underground Laboratories and Radon Abatement Systems guaranties an ideal space to study very low background physics. The LSC, with the addition of this RAS, can offer 5 orders of magnitude of cosmic rays suppression and about $1$ mBq/m$^{3}$ of $^{222}$Rn activity in air.

This reduction has been demonstrated in several types of detectors. In particular, thanks to a dedicated pipe able to deliver Rn-free air, NEXT-NEW experiment has reduced the contribution of radon to its background budget from 50\% to a negligible level. For HPGe detectors, the addition of ATEKO-air to the air flow around the detector has significantly reduced the number of counts in the peaks in the lower part of the $^{238}$U chain.

The use of RAS in the underground lab is continuously growing and will be essential in the coming months and years to further improve the radioactive background environment of experiments and technologies, particularly in the growing activities in biology, where air can not be replaced by nitrogen.


\section{ACKNOWLEDGMENTS}\label{sec:9}

This work was supported in part by the European Regional Development Fund (ERDF), corresponding to the 2013-2016 Spanish Scientific and Innovation Research Program, grants No. CPEE15-EE-3829 and No. CPEE13-4E-2646.

This work was supported in part by the Polish National Science Centre (NCN), grant No. UMO-2020/37/B/ST2/03905 and the Polish Ministry of Education and Science grant No. DIR/WK/2018/08.

Laboratorio Subterráneo de Canfranc (LSC) is a Spanish Unique Science and Technology Public Research Institution (ICTS) granted by the Spanish Ministry of Science and Innovation, The Regional Government of Aragon and the University of Zaragoza.


\bibliographystyle{plain}

\end{document}